# Soil Erosion in the United States. Present and Future (2020-2050)

**Shahab Aldin Shojaeezadeh[1], Malik Al-Wardy*[1], Mohammad Reza Nikoo[2], Mehrdad Ghorbani Mooselu[3], Mohammad Reza Alizadeh[4], Jan Franklin Adamowski[4], Hamid Moradkhani[5], Nasrin Alamdari[6], Amir H. Gandomi[7]**

1 Department of Soils, Water and Agricultural Engineering, Sultan Qaboos University, Muscat 123, Oman.

2 Department of Civil and Architectural Engineering Sultan Qaboos University, Muscat 123, Oman.

3 Department of Engineering Sciences, University of Agder 4630, Norway.

4 Department of Bioresource Engineering, McGill University, Lakeshore Road 21111, Canada.

5 Center for Complex Hydrosystems Research, Department of Civil, Construction, and Environmental Engineering, The University of Alabama, Tuscaloosa, AL 35401.

6 Department of Civil and Environmental Engineering, Resilient Infrastructure and Disaster Response (RIDER) Center, Florida A&M Univ.-Florida State Univ. College of Engineering, Tallahassee, FL 32310.

7 Faculty of Engineering and Information Technology, University of Technology Sydney, Sydney 2050, Australia.

***Malik Al-Wardy**

**Email:** mwardy@squ.edu.om

## Abstract

Soil erosion is a significant threat to the environment and long-term land management around the world. Accelerated soil erosion by human activities inflicts extreme changes in terrestrial and aquatic ecosystems, which is not fully surveyed/predicted for the present and probable future at field-scales (30-m). Here, we estimate/predict soil erosion rates by water erosion, (sheet and rill erosion), using three alternative (2.6, 4.5, and 8.5) Shared Socioeconomic Pathway and Representative Concentration Pathway (SSP-RCP) scenarios across the contiguous United States. Field Scale Soil Erosion Model (FSSLM) estimations rely on a high resolution (30-m) G2 erosion model integrated by satellite- and imagery-based estimations of land use and land cover (LULC), gauge observations of long-term precipitation, and scenarios of the Coupled Model Intercomparison Project Phase 6 (CMIP6). The baseline model (2020) estimates soil erosion rates of 2.32 Mg ha$^{-1}$ yr$^{-1}$ with current agricultural conservation practices (CPs). Future scenarios with current CPs indicate an increase between 8% to 21% under different combinations of SSP-RCP scenarios of climate and LULC changes. The soil erosion forecast for 2050 suggests that all the climate and LULC scenarios indicate either an increase in extreme events or a change in the spatial location of extremes largely from the southern to the eastern and northeastern regions of the United States.

## Significance Statement

This study employs the most recent high-resolution climate change (1-km), land use (30-m), and satellite imagery datasets (10-m) to provide a comprehensive assessment of water-driven soil erosion with a 30-m field-scale resolution in baseline (2020) and future (2050) years. Previous limitations in the literature regarding the high-resolution estimation/prediction of potential soil erosion in various landscapes are addressed in this research. We use several methodologies, from deep learning and machine learning models to satellite- and imagery-based estimations, that incorporate the impact of land use and land management strategies, such as contouring, in soil erosion estimation. Our investigation into the economic and ecological consequences of soil erosion can assist policymakers in evaluating the impacts of soil erosion under climate change and adopting best management conservation practices at the regional- to continental-scale.

## Main Text

## Introduction

Water-induced soil erosion has affected a diverse variety of natural ecosystems, agricultural and forest lands, and urban areas across the contiguous United States (CONUS) since 1930 (1, 2). Under natural conditions, a millimeter of soil takes over 25 years to form (equivalent to 0.04 mm $yr^{-1}$), whereas the erosion of agricultural soils removes soil by an average rate of 0.5 mm $yr^{-1}$ (1, 3, 4). Soil erosion estimation procedures and soil erosion management plans have been improved using advanced approaches such as a broad range of land use/land cover (LULC) cropland classes in soil erosion estimation, creating spatially distributed soil erosion maps, and defining long-term research goals for the CONUS (5–8). However, soil erosion in cropland areas increased by 3.5% between 2007 and 2017 (9–11), resulting in nutrient and organic matter loss (12), increased runoff (13), water pollution (14), productivity loss (15), eutrophication, and sedimentation (16) — with a total cost of USD 44 billion (17). Soil erosion from agricultural lands can result in a significant loss of soil organic carbon (SOC). Surface soil losses also reduce soil respiration leading to lower cumulative gaseous emissions at the eroded site — although normal emission rates occur at deposition sites — thereby lowering the soil's ability to sequester and store $CO_2$ (18, 19). In certain regions, such as croplands and deserts, soil erosion can also lead to accelerated desertification (6).

Despite the plethora of studies addressing the effect of climate change and/or LULC shifts on soil erosion (20–22), most research either focused strictly on arable agriculture and ignored LULC changes on soil erosion (23, 24) or omitted the effect of soil erosion management practices in their estimations (20–22, 25, 26). Additionally, previous research estimated and projected soil erosion for present and future circumstances using a restricted number of climate stations and a limited number of LULC classes (e.g., disregarding urban areas, 4, 22, 27). This may have introduced uncertainty into their estimations and projections. Most global and continental-scale studies of soil erosion estimation are based on the Revised Universal Soil Loss Equation (RUSLE) (28–33), which integrates climate and LULC properties to provide a spatial estimate of soil erosion (29). Former studies employed a variety of approaches and techniques to calculate RUSLE factors (4, 24, 34, 35) and develop empirical or machine learning (ML) models to project the effects of different climate scenarios (25, 36–39). These studies revealed that soil erosion decreased between 1970 and 2010, although an increasing trend in soil erosion was observed between 2010 and 2020, as well as under future climate and LULC change scenarios (1, 4, 8, 20). A recent study suggested that soil erosion in the United States could reach as high as 4.22 Mg $ha^{-1}$ $yr^{-1}$ by 2070 (4). However, estimation of soil erosion rates in the present and probable future, using various LULC scenarios and taking practice management strategies into account, have not been thoroughly explored at the field scale in the CONUS.

To provide new insights into soil erosion in the United States, we employed a Deep Learning (DL) model integrated with a Gaussian Process Regression (GPR) interpolation kernel and Random Forest (RF) ML model to estimate/project rainfall erosivity, as well as a sustainability factors. Drawing upon data from factorial combinations of three alternative (2.6, 4.5, and 8.5) Shared Socioeconomic Pathway and Representative Concentration Pathway (SSP-RCP) scenarios of CMIP6, we used thr FSSLM to calculate soil erosion at a 30-m resolution for the baseline year of 2020 and future year of 2050 under various climate/LULC scenarios. For the baseline year, we used 1971-2003 rainfall data from 3221 National Oceanic and Atmospheric Administration (NOAA) weather stations and combined Sentinel-2 and Landsat satellite products and National Agriculture Imagery Program (NAIP) imagery. CMIP6 pathways were applied to investigate present and future soil erosion trends under various soil erosion management practices (*e.g.*, alley cropping, contour farming) and LULC classes. Our research on soil erosion estimation/projection provides knowledge to policymakers and local governments across the United States to plan for soil erosion on- and off-site assessment and management strategies.

## Results

Averaged across all LULC classes, the FSSLM is estimated the soil erosion as 2.32 Mg ha$^{-1}$ yr$^{-1}$ for the baseline year 2020 (Fig. 1A). Soil erosion rates were greatest in the upstream and downstream portions of the Mississippi River, as well as in the downstream portion of the Missouri River (Fig. 1A), where the main crops were corn (*Zea mays* L.) and soybeans [*Glycine max* (L.) Merr.]. A similar pattern existed in Kansas, Oklahoma, and Texas, where current cotton (*Gossypium hirsutum* L.), sorghum [*Sorghum bicolor* (L.) Moench], and winter wheat (*Triticum aestivum* L.) production could be maintained under annual soil erosion tolerance rates but was high enough to harm environmental habitats. To manage the consequences of soil erosion, the southern regions of the United States are mainly protected by bedding, hillside ditches, and water and sediment control basins (Fig. S4B) (40). In contrast, northern croplands are primarily protected by alley cropping, contour farming, or hedgerow planting to constrain soil erosion processes within crop fields (Fig. S4A) (40). Although the eastern shores and western states, including Washington and California, had high soil erosion rates (~1.25 Mg ha$^{-1}$ yr$^{-1}$), they typically met soil erosion tolerance standards (Fig. S5B). In contrast, in the northern regions of the United States (particularly upstream of the Mississippi River), soil erosion exceeded the tolerance limit (Fig. S5B). Soil erosion was highest in croplands cultivated with corn, soybean, sorghum, and wheat, and lowest in forestland, shrublands, and grasslands (Fig. S6). Furthermore, soil erosion rates were lowest at higher elevations and in deserts and highest in alluvial and fluvial areas (Fig. S10).

Soil erosion in cultivated croplands was estimated to be 4.72 Mg ha$^{-1}$ yr$^{-1}$, which is close to the National Resource Inventory (NRI) estimate of 2017 (~5.91 Mg ha$^{-1}$ yr$^{-1}$) (refer to the Supplementary Information (SI) Appendix, Results and Discussion section). By our estimation, cultivated croplands (*e.g.*, permanent and seasonal crops) covered 17% of the United States land area in the baseline year (2020), and were responsible for 44% of soil erosion, equivalent to 0.784 Pg yr$^{-1}$. Combined cultivated and non-cultivated croplands (*e.g.*, forest, hay, pastureland, grassland, and shrubland) covering 75% of the United States lands, were responsible for 90% (~1.93 Pg yr$^{-1}$) of the soil erosion in the CONUS. With the exception of Woznicki et al., 2020 (22), most previous studies ignored soil erosion from non-cultivated croplands

Population growth and greenhouse gas emissions will radically alter LULC patterns in the United States in the future (Fig. S11). Although these patterns are highly nonlinear, on average, the expansion of urban (~66%) and cropland (~7%) areas from 2020 to 2050 will alter soil erosion rates and threaten new lands (*e.g.*, the northeast United States) by introducing a new pattern of soil erosion that is monotonically increasing across the CONUS (Fig. S14). To assess the impact of LULC changes, we projected soil erosion until 2050 using combinations of Shared Socioeconomic Pathway and Representative Concentration Pathway (SSP-RCP) scenarios, namely SSP1-RCP 2.6, SSP2-RCP 4.5, and SSP5-RCP 8.5 (Fig. 1B-D) and considering the baseline rainfall erosivity.

The SSP1-RCP 2.6 scenario is a shift in world development toward a more sustainable path, raising global temperature from 1.3°C to 2.4°C by 2100. When compared with baseline soil erosion (2.32 Mg $ha^{-1}$ $yr^{-1}$), the SSP1-RCP 2.6 scenario showed a total increase of 4.6% in soil erosion by 2050, reaching 2.43 Mg $ha^{-1}$ $yr^{-1}$. This increase in soil erosion is attributable to a rapid expansion in agricultural (~ 6%) and urban areas (~ 55%), as well as a reduction in forested lands (~-5%) (Fig. S14). The slight increases were due to agricultural expansion in the cropland states (e.g., Iowa, Illinois, and Missouri). Although these regions featured moderate baseline rainfall erosivity (~2300 MJ $mm^{-1}$ $ha^{-1}$ $yr^{-1}$), they contributed 13% to the nation's overall soil erosion at baseline, which increases to 15% by 2050 under the SSP1-RCP 2.6 scenario. To continue the present path, the SSP2-RCP 4.5 follows social, economic, and technological trends that do not shift from historical patterns; however, greenhouse gas (GHG) emissions are low and constant till 2050 and then decrease. According to this scenario, overall soil erosion could increase by 7.8% in the CONUS (reaching 2.50 Mg $ha^{-1}$ $yr^{-1}$) (Fig. 1C), mainly due to intensive agricultural expansion (~ 15.5%) and deforestation (~ -8%) (Fig. S14). In comparison, under the SSP5-RCP 8.5 scenario, fossil fuel-based development persists, there are no environmental limitations, and 2075 GHG emissions are 3-fold greater than those in 2020. Although the cropland expansion (~13%) is lower than that under SSP2-RCP 4.5 scenario, the intensive deforestation (~-12%) and increased urbanization (~88%), lead to a 7.4% increase in soil erosion, mostly in the southeastern and eastern portions of the lands which were previously forest or shrublands at 2020 (Fig. 1D).

Climate change has a significant impact on rainfall frequency and intensity. Based on our analysis, from 1979 to 2013, rainfall erosivity calculated from 15-minute rainfall intervals drawn from 3221 NOAA stations in the CONUS (Fig. S2), increased on average by 2.00 to 2.14 GJ $mm^{-1}$ $ha^{-1}$ $yr^{-1}$. The eastern United States showed the greatest rainfall erosivity, exceeding 13 GJ $mm^{-1}$ $ha^{-1}$ $yr^{-1}$ in the state of Louisiana near the Gulf of Mexico (Fig. 2A). The western regions showed the lowest values, with a minimum of 13 MJ $mm^{-1}$ $ha^{-1}$ $yr^{-1}$ in Arizona and Nevada. To investigate the future effect of rainfall erosivity on soil erosion, we projected rainfall erosivity for the year 2050 using a DL model with an integrated GPR kernel, and drawing upon three alternative (2.6, 4.5, and 8.5) Shared Socioeconomic Pathway and Representative Concentration Pathway (SSP-RCP) scenarios of CMIP6 (Fig. 2B-D). On a continental scale, average rainfall erosivity will increase by 24% (SSP1-RCP 2.6), 27% (SSP2-RCP 4.5), and 31% (SSP5-RCP 8.5), with increasing trends in the eastern and southeastern United States and decreasing trends in the northwestern, north, and southern regions (Fig. 2B-D). Rainfall erosivity as a major driver will therefore change dramatically by 2050 regardless of the LULC scenarios. Further investigation revealed that lower (*vs.* higher) elevations would be subject to greater rainfall erosivity (Fig. S10).

We combined the effect of projected rainfall erosivity and LULC changes for 2050 for the best (SSP1-RCP 2.6) and the worst ( SSP5-RCP 8.5) scenarios (Figs. 3A, 3B, respectively). The model indicated that soil erosion rates could increase by 8% (SSP1-RCP 2.6), 15% (SSP2-RCP 4.5), and 21% (SSP5-RCP 8.5), under climate/LULC scenarios. The SSP1-RCP 2.6 climate/LULC scenario showed the smallest and the least likely increase in soil erosion by 2050. However, changing hydrological cycles may accelerate the rising soil erosion rate in northern and southeastern regions (~twice the baseline), where most LULC scenarios suggest cropland expansion relative to the 2050 baseline. The SSP1-RCP 2.6 is optimistic and can be interpreted as the climate scenarios' lower confidence interval. In contrast, SSP2-RCP 4.5 is known to represent the conditions most likely to occur, and as an intermediate scenario, would see an increase in soil erosion. Accordingly, the SSP2-RCP 4.5 scenario was representative of the average increase in soil erosion by 2050. The SSP5-RCP 8.5 climate change scenario can be defined as the upper confidence interval of soil erosion projections which, although extremely unlikely, can be considered as the worst-case scenario in which emissions continue to rise throughout the twenty-first century.

Although the increase in soil erosion might at first appear to be excessive, our estimation of the average CONUS soil erosion rate in 2050 (~2.88 Mg $ha^{--1}$ $yr^{-1}$) is more reasonable than the estimation for a similar scenario (SSP5-RCP 8.5) made by Borrelli et al. (4), which implemented the LULC scenarios of the Coupled Model Intercomparison Project Phase 6 (CMIP6) for LULC scenarios and CMIP5 for climate changes in 2070. They suggested that under the worst-case scenario, soil erosion would increase by 82% (reaching 4.22 Mg $ha^{-1}$ $yr^{-1}$) by 2070 (4). While the timeline differences between our study and that of Borrelli et al. (4) may lead to some discrepancies in comparison, there are some other influential differences between these two studies. Such differences include: LULC geospatial resolution (30-m in this study versus 5.5-km in (4)), rainfall erosivity projection (in southern regions, we concluded a decrease, whereas (4) deduced an increase), climate and LULC scenarios (CMIP6), satellite-based estimations, non-stationary models (DL and ML models). In addition, Borrelli et al. (4) ignored management practices in their estimation which, all told, affect the results and have led the present authors to believe this study to be more accurate with lower uncertainties compared to Borrelli et al.'s projections for 2070 (4). Notably, Borrelli et al. (4) believed that using various management practices broadened the uncertainty range in soil erosion estimation. As a result, estimated soil erosion extremes in the future ranged from as low as 2.56 Mg $ha^{-1}$ $yr^{-1}$ (Fig. 3A) to as high as 2.88 Mg $ha^{-1}$ $yr^{-1}$ (Fig. 3B). The FSSLM results indicated that changes in rainfall erosivity had more significant influences on future soil erosion than changes in LULC. Further analysis revealed that the eastern and southeastern United States would experience a tremendous increase (more than twice) in soil erosion within both the worst (SSP5-RCP 8.5) and the best (SSP1-RCP 2.6) case scenarios. While the western United States showed some deviations from the baseline, soil erosion in the northern regions would be reduced (as rainfall erosivity decreased, Fig. 2B-D) by 2050.

Additionally, we analyzed soil erosion, soil fertilizer consumption and expense (41, 42), soil respiration (43), and the interactive effects of these factors across the CONUS. Fig. 4 depicts the change in the rate of soil erosion in each state between 1997 (left) and 2012 (right). It shows increases in Iowa, Missouri, and Wisconsin and decreases in North Carolina, Tennessee, and Maryland. Soil erosion also increased fertilizer consumption (either nitrogen or phosphorus) in agricultural states like Iowa, Nebraska, Texas, Minnesota, and Illinois, which have the highest rates of agricultural production (using phosphorus fertilizers), the lowest soil respiration, and the highest rates of soil erosion. In contrast, Wyoming, Utah, and Maine had lower soil erosion, higher soil respiration, and lower fertilizer consumption and expenditures due primarily to large expanses of forested lands. Because soil erosion reduces soil respiration and fertility, farmers are forced to use large quantities of fertilizer to restore soil nutrients, which can cause environmental problems such as contaminants leaching into rivers, groundwater, or polluting food chain systems. While soil fertilizer inputs should be increased to maintain soil fertility for crop production as soil erosion increases, this feedback loop has the potential to exacerbate environmental issues, demonstrating the importance of soil erosion protection, particularly in agricultural states. In general, the rate of soil respiration and fertilizer costs in the CONUS has increased over time, particularly in the northern states; however, as soil erosion decreased, soil respiration increased, but soil fertilizer consumption and costs have risen dramatically. Therefore, given the increase in soil erosion by 2050, on- and off-site consequences could be either increased or shifted across the United States.

## Discussion

Originally developed for agricultural lands, the RUSLE is commonly used to estimate water-driven soil erosion; however, the G2 erosion model addresses some limitations on soil erosion estimation for different types of sparse vegetation (e.g., grassland and shrublands). While recent studies have adopted the RUSLE and included its cover management factor in estimating soil erosion for non-agricultural LULC types (*e.g.*, forested and urban areas), most studies omitted management practices in their estimations. We assessed the impact of various National Resource Conservation Service (NRCS) practices used across the CONUS on soil erosion in the baseline year 2012 (Fig. S4A-B, Table S4). Accordingly, by considering these conservation management practices and estimating available practices by satellite and imagery databases in 2020, soil erosion was estimated. Also, by holding NRCS practices constant over time to project soil erosion rates in 2050, our findings differed from previous local and global soil erosion estimations, both in the present and future scenarios. Relative to the FSSLM's baseline results (2020), our results showed an increase of approximately 0.23 Mg $ha^{-1}$ $yr^{-1}$ in total soil erosion compared to Woznicki et al., 2020 (22), an increase of approximately 0.1 Mg $ha^{-1}$ $yr^{-1}$ compared to Borrelli et al., 2020 (4), and a decrease of approximately 0.36 Mg $ha^{-1}$ $yr^{-1}$ compared to Borrelli et al, 2017 (44), clearly demonstrating the importance of various practices methods on soil erosion controlling and soil erosion estimations, although also showing estimated soil erosion rates comparable to previous studies.

In recent decades, climate change has had various effects on soil erosion (Fig. 4). Between 1982 and 2012, after implementing various USDA soil erosion control methods and conservation practices, soil erosion was reduced by ~34% (9). By 2017, however, due to increases in precipitation intensity and frequency (37%) and an increase in agricultural land expansion (2.5%), soil erosion further increased by ~3.5% (to 6 Mg $ha^{-1}$ $yr^{-1}$) in croplands between 2012 and 2017 (10, 11). Our estimation of rainfall erosivity in the CONUS at a sub-km resolution between 1979 and 2013 indicated that rainfall erosivity had increased while rainfall volume and duration decreased (Fig. S11). While previous estimations of rainfall erosivity exist across the CONUS [*e.g.*, isoerodents (37, 45) or Param-elevation Regressions on Independent Slopes Model (PRISM) estimates (46)], the estimation period (1970-2000) and spatial resolution (4 km) were insufficient to account for the spatial distribution of climatic factors on soil erosion. The recent Global Rainfall Erosivity Database (GloREDa) (25), satellite-based estimations of rainfall erosivity (47), and updated isoerodent maps of the United States (48) were plagued with spatiotemporal uncertainties arising from the periods of estimation, the number of climate stations, 30-min interval assumptions, and stationary methods (*e.g.,* kriging methods) employed in interpolation (49). Although the rainfall erosivity pattern is non-stationary, most recent studies have projected rainfall erosivity using either empirical or stationary ML models (35, 38, 39, 48). Additionally, various patterns for rainfall erosivity changes have been proposed, but most studies, except for Borrelli et al., 2020 (4), suggested that in the future, rainfall erosivity would increase in the eastern and southeastern United States and decrease in the southern regions (24, 26, 50). For future climate change scenarios, the DL model integrated with GPR interpolation kernel employed to estimate rainfall erosivity found patterns of rainfall erosivity similar to those of previous studies (24, 26, 50).

## Conclusions

The Field Scale Soil Loss Model (FSSLM) estimated soil erosion in 121 LULC classes using satellite imagery databases under baseline (2020) conditions and 20 LULC classes in 2050 at the field scale (30-mmm resolution). Our findings indicated that soil erosion rates would not only increase overall but, in some scenarios, would be shifted spatially from southern and northern regions to new hotspots in the eastern and southeastern United States. While climate factors have the most significant influence on increasing soil erosion, LULC shifts are also involved. When current soil erosion practices are combined with the worst-case future scenario (SSP5-RCP 8.5), the greatest increase in soil erosion is 21% higher than the baseline estimation (~2.32 Mg $ha^{-1}$ $yr^{-1}$). The FSSLM can address county-level economic and environmental consequences of soil erosion in terms of soil fertilizer consumption and soil respiration. Our findings contribute to a better understanding of soil erosion and its consequences across the CONUS, and provide some critical insights into the efficacy of soil erosion control strategies based on current practices and future projections, previously indicated to have serious impacts on soil erosion estimation (51). The FSSLM adopts public and reproducible data for the contiguous United States to provide a robust estimation of soil erosion that can be upgraded when new data becomes available. The FSSLM is proposed to enable policymakers and scientists to evaluate current practices for present and future climate and LULC changes with a new non-stationary DL model combined with satellite-based data to interpret agricultural lands modifications.

# Materials and Methods

## Datasets

The Field Scale Soil Loss Model (FSSLM) was developed based on the G2 erosion model and designed to estimate water-driven soil erosion in current and future periods by combining field measurements with remote sensing datasets, and producing a new projection based on the latest IPCC scenarios (CMIP6) for land use and climate change. The model combined several satellite, cropland, and climate data layers, including USDA's CropScape, National Agriculture Imagery Program (NAIP), Sentinel-2 imagery, Landsat Net Primary Production (NPP) and Gross Primary Production (GPP), Land-Use Harmonization version 2 (LUH2), land use projection using cellular automata based on CMIP6 scenarios (52), and A Climate Adaptation Conservation Planning Database for North America (AdaptWest) (53) climate change models. These datasets were chosen to provide standard output for United States government authorities as well as to enable an accessible and reproducible knowledge exchange for adjacent disciplines in future implementation and treatment at a field scale. Further details are available in the SI Appendix Materials and Methods section.

## Model Limitations

With recent advancements in rainfall erosivity estimation, the FSSLM provides a new opportunity to estimate soil erosion at a high resolution for examination of the Sentinel-2 imagery dataset at each pixel. However, this model inherits the limitations of RUSLE and 'G2', including: (*i*) it is semi-empirical and (*ii*) it is data-driven, and (*iii*) it estimates only sheet and rill erosion. Although these constraints may introduce uncertainty into the modeling, no other method or model can provide such information. The FSSLM framework limitations are detailed in the Model limitations section of the SI Appendix.

## Model Evaluation

We evaluated our estimations with National Resource Inventory (NRI) field measurements in 2017 and the Daily Erosion Project (DEP) model. The G2 erosion model's sustainability factor was evaluated individually by NRI estimation for each state and crop type. This method investigated the sustainability factor for various crop types such as corn, soybean, and wheat and provided information for each crop under each county's current (2020) management practices (40). Cover sustainability factors for other land use types were obtained from the literature (refer to the SI Appendix, Results and Discussion section).

We calculated rainfall erosivity using 1970 to 2013 NOAA data from 3221 stations across the CONUS (54). A DL model integrated with a GPR kernel was used to interpolate rainfall erosivity between 1970 and 2013 and project it using the AdaptWest climate dataset. Thirteen different General Climate Models (GCMs) were used to project rainfall erosivity, and, based on previous studies (55), the ensemble average was chosen to reduce uncertainties. We then compared our findings to previous rainfall erosivity models and discovered some discrepancies (refer to the SI Appendix, Results and Discussion section).

Despite the various DL, ML, satellite, and imagery models and evaluation methods used to predict and project soil erosion, some geographical and scale uncertainties, as well as climate and LULC projections, limited the accuracy of actual estimation in the future. The FSSLM attempted to reduce these uncertainties, particularly using high resolution (the best available) climate and LULC datasets and recent CMIP6s scenarios, to contribute to a better understanding of soil erosion in the United States. The results reveal new hotspots of potential future soil erosion to authorities and policymakers and deliver managerial information. More detailed analyses are available in the SI Appendix under the Results and Discussion section.

**Data Availability**

The data presented in Fig. 3 are available at: https://zenodo.org/record/6413769, USDA's Cropscape database is available at: https://nassgeodata.gmu.edu/CropScape, National Agriculture Imagery Program (NAIP) dataset of each county is available through earth explorer: https://earthexplorer.usgs.gov, Sentinel-2 imageries can be accessed from sentinel-hub: https://www.sentinel-hub.com, Landsat Net Primary Production (NPP) and Gross Primary Production (GPP) are available through the University of Minnesota portal: http://files.ntsg.umt.edu/data/Landsat_Productivity, Land-Use Harmonization version 2 (LUH2) is available at: https://luh.umd.edu, land use projection using cellular automata based on CMIP6 scenarios is deposited in: https://zenodo.org/record/4584775, and A Climate Adaptation Conservation Planning Database for North America (AdaptWest) is available at: https://adaptwest.databasin.org.

**Acknowledgments**

The authors would like to acknowledge the financial support provided by the Ministry of Environment and Climate Affairs (Environment Authority) of the Sultanate of Oman, under the Project for Mapping, Monitoring and Mitigation of Land Degradation in Oman (CR/AGR/SWAE/13/02), awarded to the Sultan Qaboos University, Oman.

**Figures and Tables**

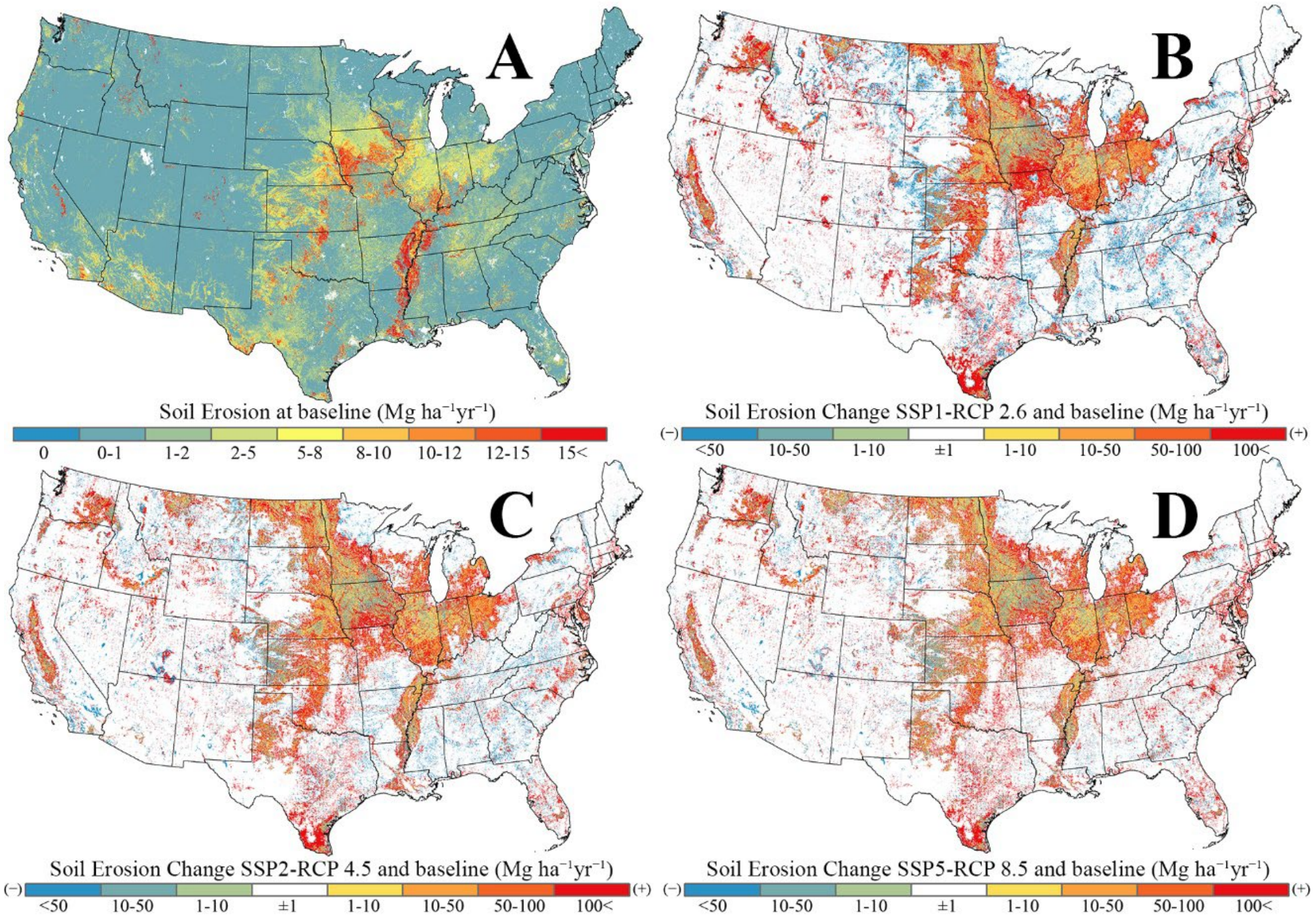

**Figure 1 (A) Soil erosion estimates in baseline, soil erosion changes between baseline and scenario of (B) SSP1-RCP2.6, (C) SSP2-RCP4.5, (D) SSP5-RCP8.5 without rainfall erosivity projection (rainfall erosivity of baseline) at 30-m resolution.**

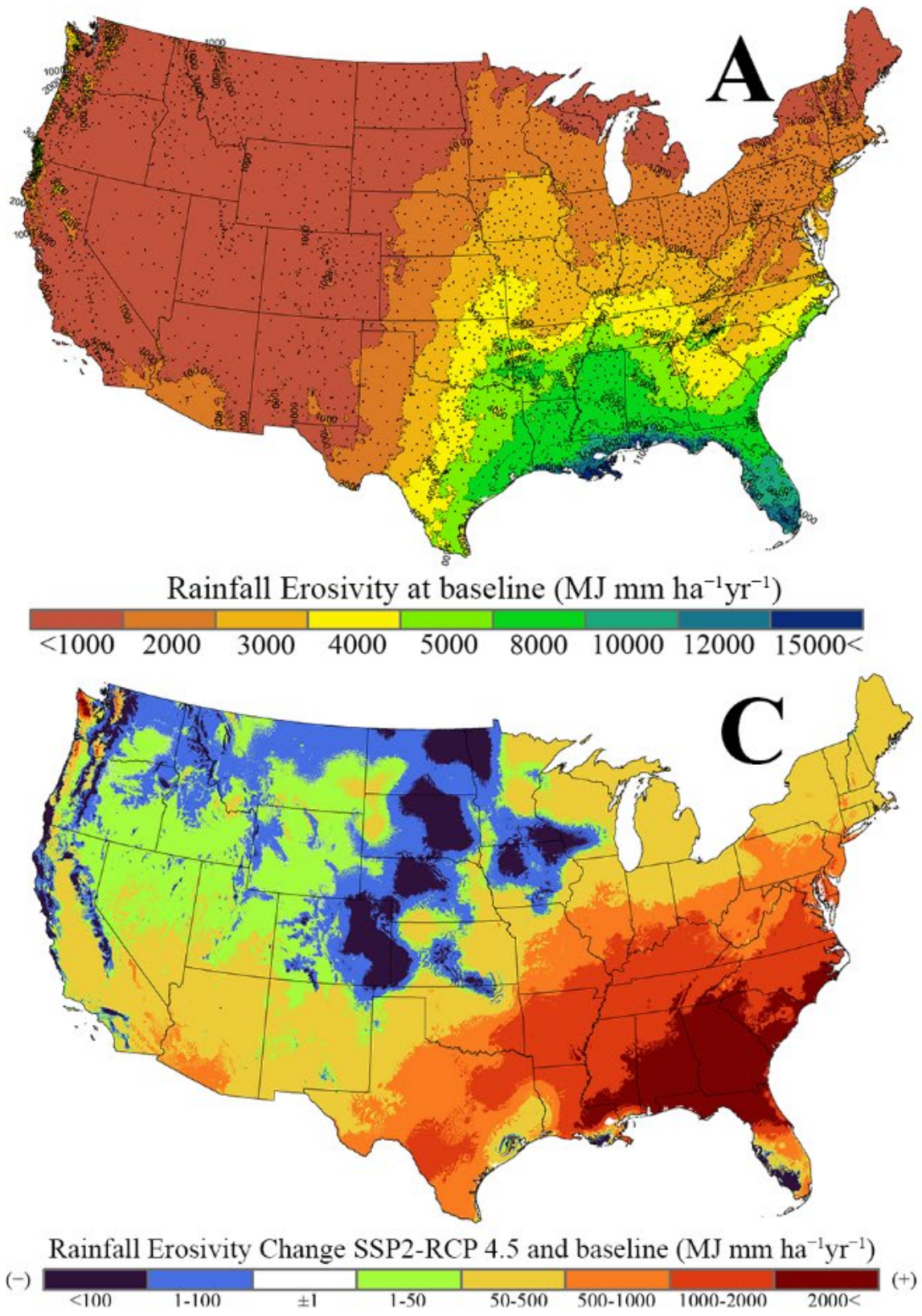

**Figure 2 Annual rainfall erosivity estimates (MJ mm$^{-1}$ ha$^{-1}$ yr$^{-1}$) at 30 arc-seconds (A) values between 1979 and 2013 estimated by integrated DL and GPR model, changes in rainfall erosivity between baseline and scenario of, (B) SSP1-RCP2.6, (C) SSP2-RCP4.5, and (D) SSP5-RCP8.5, based on average values of 13 GCMs of the AdaptWest dataset version 1.1 (Table S3).**

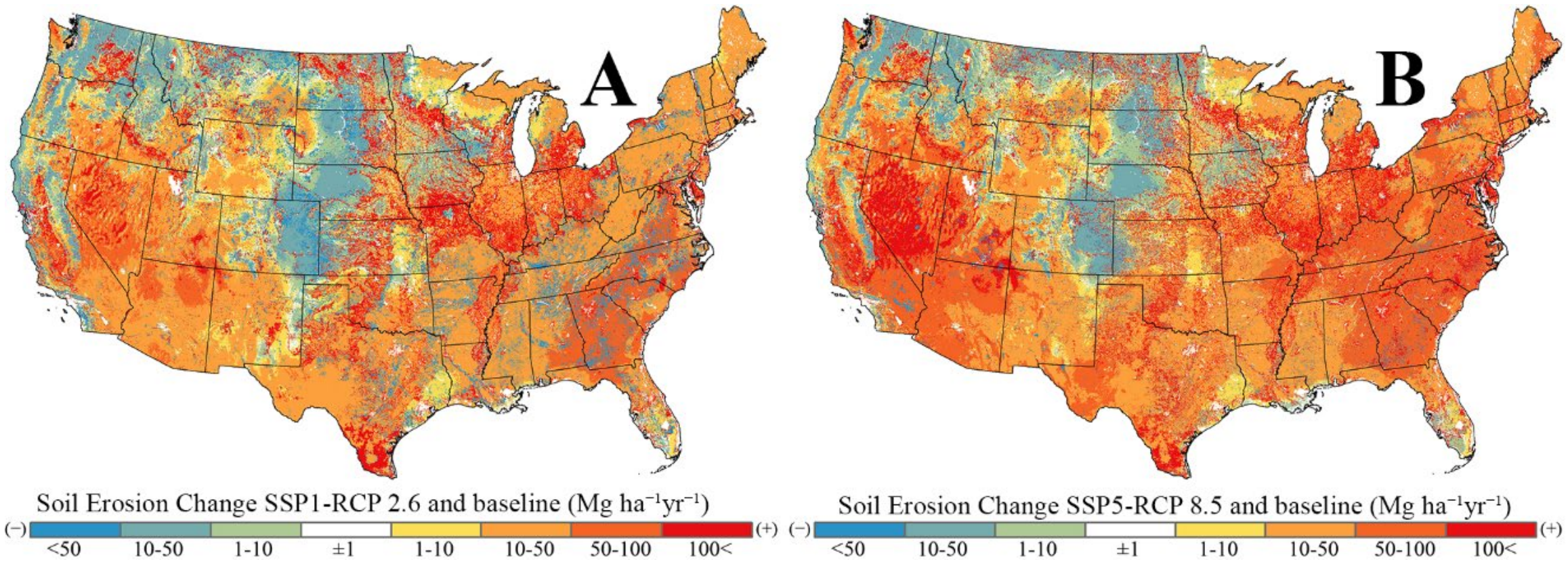

**Figure 3 Soil erosion changes between baseline and scenarios of (A) SSP1-RCP2.6, and (B) SSP5-RCP8.5 at 30-m resolution.**

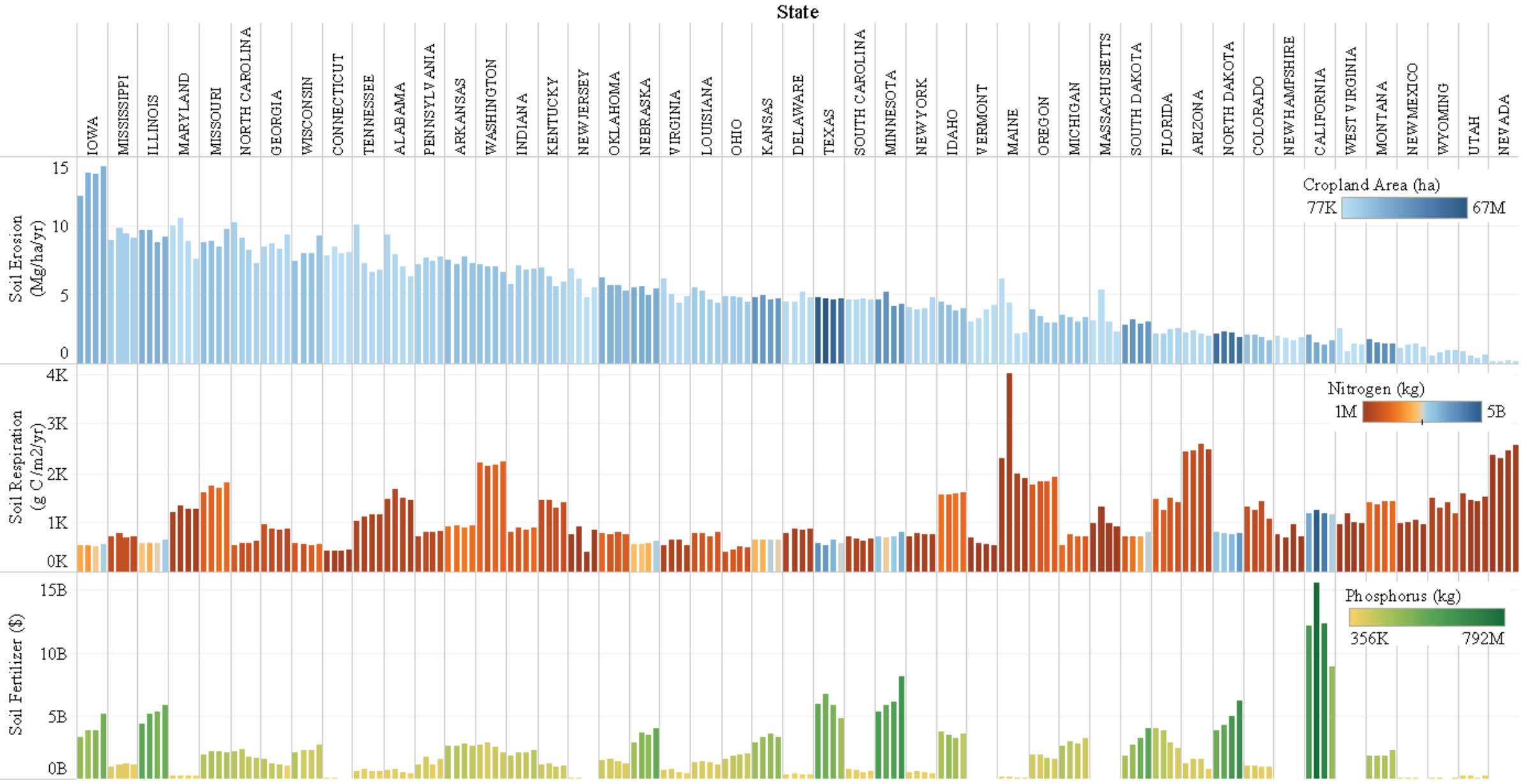

**Figure 4 (Top) Soil erosion (Mg $ha^{-1}$ $yr^{-1}$), (Middle) soil respiration (g $cm^{-2}$ $yr^{-1}$), and (Bottom) soil fertilizer ($) in each state from (Left) 1997 to (Right) 2012 in 5-year intervals. Color pallets refer to, (Top) Cropland area (ha), (Middle) Nitrogen (kg), (Bottom) Phosphorus (kg). Soil fertilizer consumption expenses ($) are normalized to 2011.**

**Supplementary Information for Soil Erosion in the United States. Present and Future (2020-2050)**

**Shahab Aldin Shojaeezadeh[1], Malik Al-Wardy*[1], Mohammad Reza Nikoo[2], Mehrdad Ghorbani Mooselu[3], Mohammad Reza Alizadeh[4], Jan Franklin Adamowski[4], Hamid Moradkhani[5], Nasrin Alamdari[6], Amir H. Gandomi[7]**

1 Department of Soils, Water and Agricultural Engineering, Sultan Qaboos University, Muscat 123, Oman.

2 Department of Civil and Architectural Engineering Sultan Qaboos University, Muscat 123, Oman.

3 Department of Engineering Sciences, University of Agder 4630, Norway.

4 Department of Bioresource Engineering, McGill University, Lakeshore Road 21111, Canada.

5 Center for Complex Hydrosystems Research, Department of Civil, Construction, and Environmental Engineering, The University of Alabama, Tuscaloosa, AL 35401.

6 Department of Civil and Environmental Engineering, Resilient Infrastructure and Disaster Response (RIDER) Center, Florida A&M Univ.-Florida State Univ. College of Engineering, Tallahassee, FL 32310.

7 Faculty of Engineering and Information Technology, University of Technology Sydney, Sydney 2050, Australia.

***Malik Al-Wardy**

**Email:** mwardy@squ.edu.om

# SI Materials and Methods

# 1. Field Scale Soil Loss Model (FSSLM)

## 1.1. G2 Erosion Model

The modified Revised Universal Soil Loss Equation (RUSLE), known as the G2 erosion model, empirically estimates climate and land-use variables (*e.g.*, rainfall erosivity, vegetation retention, and landscape effect) over time, while other random variables (*e.g.*, soil erodibility, and terrain influence) are assumed to be deterministic. The G2 erosion model predicts monthly or annual soil loss values and considers rainfall erosivity as having a direct effect (the numerator in Eq. 1), whereas the erosion-controlling factors of vegetation retention and landscape effect have an inverse (the denominator in Eq. 1) effect. Other factors only increase or decrease the magnitude of related factors, *e.g.*, raindrop kinetic energy and the rate of associated runoff. While it has some limitations, this model is well-known, widely-used, and generally deemed the best available method to estimate soil erosion based on a detachment-limited model (1). Soil erosion prediction based on the G2 erosion model represents the result of sheet and rill erosion processes, which are defined by the following equation:

$$A = \frac{R \cdot K \cdot T}{V \cdot L} \tag{1}$$

where, A is the annual soil erosion (Mg $ha^{-1}$ $yr^{-1}$); R, the rainfall erosivity factor (MJ mm $h^{-1}$ $ha^{-1}$ $yr^{-1}$); V, the vegetation retention factor (similar to cover management in the RUSLE equation, dimensionless); K, the soil erodibility factor (Mg $ha^{-1}$ $MJ^{-1}$ $mm^{-1}$); T, topographic effect (dimensionless); and L, natural or human-induced anti-erosion factors (similar to practice management in RUSLE equation, dimensionless). The sustainability factor (*S)* is the inverse of the product of V and L ($S = \frac{1}{V \times L}$). Here, we integrated the G2 erosion model with the proposed Field Scale Soil Loss Model (FSSLM) to estimate soil erosion.

## 1.2. Model Platform

The FSSLM is a platform based on the G2 erosion model that intrinsically multiplies various factors, including a driving force (rainfall erosivity factor, R), a soil resistance (K) term, along with the anthropogenic effects factor (S=V×L), and a topographical contribution (T) to estimate soil erosion. This model is a modified version of the well-known RUSLE equation, widely-used in regional and global studies (2–7) to estimate soil loss and soil erosion. The G2 erosion model is a GIS-based model that estimates soil erosion in each pixel isolated from other pixels in a plausible format (raster) to be easily regionalized in/for local scales. Therefore, a 30-m (field scale resolution) is sufficient to consider the raster assumption as the representative scale of the model. Except for ice sheet erosion (ICEP), the G2 erosion factors in this platform are estimated individually to assess soil erosion on a global scale if the required dataset is available.

The FSSLM estimates soil erosion based on the above-cited factors, expressing it as a mass of soil loss per unit area per unit of time (Mg $ha^{-1}$ $yr^{-1}$) at a 30-m GIS modeling resolution. Estimated factors were thus more precise and accurate when dealing with uncertainty. The flowchart of the FSSLM to estimate soil erosion under various climates (Climate section), land use and land cover (Sustainability section), soil resistance (Soil section), and topographical confirmation of field effects (Terrain section), is presented in Fig. S1.

The R-factor measures the kinetic energy of rainfall when it drops on bare or covered soil. To estimate this factor, we used high-resolution (15-min interval) gauge-based observations from 3221 active National Oceanic and Atmospheric Administration (NOAA) precipitation stations between 1971 and 2013. While long-term observations (~20 years) could be sufficient to allow reliable estimations of the rainfall erosivity factor, 50 years of observations is deemed ideal (8), a period similar to that used in the present study (~44 years). The detailed procedure of the estimation, projection, and interpolation of this factor is outlined in section 2, Climate.

The K-factor measures the inherent resistance of soil to erosion. To estimate this factor, we used a high resolution (30-m) Probabilistic Remapping of Soil Survey Geographic Database (POLARIS), which is a modified version of the Soil Survey Geographic Database (SSURGGO), based on soil observations. The detailed procedure for the estimation of soil erodibility is described in section 3, Soil.

The S-factor represents the sustainability of the anthropogenic effects on soil erosion control. This factor combines the effect of cover and management practices in the RUSLE equation. We used satellite and airborne imageries to estimate this factor at a 10-day resolution (~Sentinel revisit time; 5-10 days). This is the most complex factor requiring estimation in the G2 erosion model, and is explained in detail in section 4, Sustainability.

The T-factor measures the effect of slope length and steepness of hillslopes on soil erosion rates. We used the National Elevation Dataset (NED) at a 10-m resolution to estimate the slope and aspect of terrain along with the MERIT database's flow accumulation at a 90-m resolution to calculate the terrain influence on soil erosion. The detailed procedure for estimating the terrain's influence is described in section 5, Terrain.

# 2. Climate

## 2.1. Methodology

Estimating and projecting rainfall erosivity is a challenging endeavor. Generally, the G2 erosion model follows the RUSLE version 2 equation to estimate rainfall erosivity as follows:

$$EI_{30} = I_{30} \sum_{i=1}^{i=N} e_i v_i \tag{2}$$

where $e_i$ is the unit rainfall energy (MJ ha$^{-1}$ mm$^{-1}$); $v_i$ the rate of rainfall (mm h$^{-1}$); and $I_{30}$ the maximum rainfall intensity over 30 minutes. Since the rainfall intensity changes over time, the rainfall energy is calculated for each particular event $(i)$ as follows (8):

$$e_i = 0.29\left[1 - 0.72e^{-0.08i}\right] \tag{3}$$

Thus, Eqs 1 and 2 serve to calculate rainfall erosivity for every single event, and the average annual rainfall erosivity (MJ mm ha$^{-1}$ h$^{-1}$ yr$^{-1}$) is estimated as:

$$R = \frac{\sum_{j=1}^{j=n} \sum_{m=1}^{m=k_j} EI_{30_m}}{n} \tag{4}$$

where $n$ is the number of years; $k_j$ the number of events per year $j$; and $m$ is the index of every single event and corresponding rainfall erosivity.

## 2.2. Present Rainfall Erosivity Assessment

We estimated rainfall erosivity at 3221 active precipitation stations belonging to the NOAA using Rainfall Intensity Summarization Tool (RIST) software (Fig. S2). These stations documented rainfall values between 1970 or 1971 and 2013 and were spread across the contiguous United States, especially near the western shores and eastern temperate forests. Estimated rainfall erosivity at each station was used in a DL model integrated with a GPR kernel to interpolate rainfall erosivity across the contiguous United States. We used Google's TensorFlow Keras model to construct a DL model, employing a sequential group of layers consisting of two Dense layers with 100 output units and a 'relu' activation layer that was configured with an 'L1' regularizer. The final DL layer was a Dense layer with one output unit (~rainfall erosivity) and 'linear' activation. We also batch normalized inputs and used a Dropout layer with the rate of 0.1 to prevent overfitting. The final layer of the DL, combined with the GPR interpolation layer, formed a GPR model with a Rational Quadratic kernel to interpolate rainfall erosivity. We used the AdaptWest dataset as a covariate (inputs) to train the DL model (9). Various climate factors such as monthly precipitation (PPT), Hargreave's reference evaporation ($E_{ref}$), mean annual relative humidity (%) (RH), and precipitation as snow (mm) (PAS) were supplied as input in each station to train the DL model. Through this process, 70% of stations were used as training data, 10% for validation, and 20% for testing. These factors, from between 1970 and 2010, served as climate normals to construct the DL model and prepare the model to project rainfall erosivity. A linear regression between calculated and estimated/interpolated rainfall erosivity illustrates the model capability's to precisely predict rainfall erosivity (Fig. S12A).

## 2.3. Future Rainfall Erosivity Projection

Previous studies have developed rainfall erosivity maps for the future (10–12). However, their assumptions resulted in some uncertainty in their predictions. For example, most studies employed the Renard and Freimund equations to project rainfall erosivity based on monthly or annual precipitation. Such non-linear empirical equations should be revised to include other factors such as snow or humidity. Some recent studies (13, 14) have tried to consider temperature variables and other climatic terms through the GPR ML model. This simple assumption of isotropic interpolation made some inconsistencies with our study, although the overall trends should be comparable. Therefore, we projected rainfall erosivity using three alternative (2.6, 4.5, and 8.5) Shared Socioeconomic Pathway and Representative Concentration Pathway (SSP-RCP) scenarios of CMIP6 employing the constructed DL model, which is for regression and interpolation. We used an ensemble of 13 Global Climate Models (GCMs) to consider uncertainty in predicting future climate variables. These GCMs were selected to provide benchmarks for various dependency structures between the GCMs employed in a previous study (15). The list of GCMs is summarized in Table S3.

The rationale behind rainfall erosivity projection using covariates of AdaptWest climate factors is associated with the assumption that the combination of these factors is similar to climate normals in the present. It is assumed that the relations between rainfall intensity and rainfall erosivity, and combinations of climate factors are similar in different latitudes and years. It enables us to consider these relations exhibited in the future and project rainfall erosivity for future climates. However, unlike previous studies that considered temperature in rainfall erosivity estimation, we believe that correlating climate factors such as temperature with rainfall erosivity may impose uncertainty on rainfall erosivity projections. Therefore, to eliminate the effect of temperature rising/falling in higher latitudes that may increase/decrease rainfall erosivity incorrectly, we used climate factors such as precipitation in the form of snow or humidity that are believed to be more strongly correlated with rainfall intensity. It is believed that increases in temperature will increase evaporation, which in turn, increases overall precipitation. Moreover, climate change may cause shifts in wind patterns and ocean currents. Hence, we believe that by using reference evaporation and relative humidity, we could better consider the effect of rising temperature on precipitation in the future.

# 3. Soil

## 3.1. Methodology

Soil erodibility is an intrinsic factor that represents the natural resistance of soil to erosion driven by rainfall-runoff impacts. It depends on observations and should be estimated by various soil profile factors such as soil organic carbon, clay, sand, and silt content, along with soil permeability and other soil texture properties (16). The G2 erosion model follows an empirical equation proposed in RUSLE v2 (17):

$$K = 0.1317 \cdot \left[2.1 \times 10^{-4} \cdot M^{1.14} \cdot (12 - SOM) + 3.25 \cdot (s - 2) + \frac{2.5 \cdot (p-3)}{100}\right] \quad (5)$$

where $M$ is the textural factor and defined as $M = (m_{silt} + m_{sand})(100 - m_c)$, in which $m_c$ is the percent of clay, $m_{sand}$ is the percent of sand, and $m_{silt}$ is the percent of silt. Also, $SOM$ indicates the percentage of soil organic matter; $s$ is soil class structure, and $p$ represents the permeability class.

## 3.1. Soil Erodibility Assessment

Various soil databases are available globally (*e.g.*, SoilGrids 250-m) and locally (*e.g.*, POLARIS 30-m, SSURGGO 5-m) that provide clay, silt, sand, and SOM contents. However, permeability classes and soil class texture must be calculated using other datasets (*e.g.*, HiHydroSoil v2 database).

The POLARIS dataset was used to derive the contents of clay, sand, silt, SOM (scaling soil organic carbon to SOM using a 1.72 factor as suggested by (18)). Then, the bulk density was employed to estimate the packing density of the soil to reflect the effect of subsoil compactness in texture classification (19) at a resolution of 30-m for the first three layers' depth ranges (0-0.30 m) (20). The effect of SOM on texture classification (when SOM is 20 %, and clay content is 0%, or when SOM is 30 %, and clay content is 50 %) was also added to texture classification (21) to calculate soil structure classes at a 30-m resolution (16). Then, permeability classes were calculated using the hydrologic group of soil from the HiHydroSoil v2 dataset (22) and saturated hydraulic conductivity ($k_{sat}$) from the POLARIS database, which was modified by the volumetric fraction of coarse fragments from the SoilGrids database. This was mainly to modify $k_{sat}$ and eliminate the effect of rock fragments (stoniness) on soil erodibility estimation (23). Finally, the soil erodibility factor was estimated assuming POLARIS variability at a 250-m resolution of SoilGrids and HiHydroSoil v2 databases based on raster calculation of GIS in 30-m (Fig. S7A).

Soil erodibility is also available through the SSURGO database. However, some limitations of this database include: (*i*) unmapped areas, which are gap-filled using survey data from the surrounding regions and Soil Survey and supersede the State Soil Geographic (STATSGO) v2, (*ii*) discontinuities at state political boundaries, (*iii*) polygon-based soil properties, and (*iv*) a restricted range of soil erodibility values (14 classes), which is insufficient to present real values at a 30-m resolution, such as this study. How the present study addressed these inconsistencies is shown in (Fig. S7A-B).

# 4. Sustainability

## 4.1. Methodology

### 4.1.1. Vegetation retention (V)

The RUSLE equation's cover management spin-off is the vegetation retention factor, which is a more logical method of looking at the effect of natural or agricultural vegetation on erosion prevention. This factor identifies three types of effects: (*i*) the canopy effect (also suggested to be considered in the RUSLE2 equation as the effect of crown density and closure degree above the surface, (24), (*ii*) small plants and residues effect as natural protection on the surface (25), and (*iii*) SOM and soil structure enhancement beneath the surface, which can express the effect of crop rotations on erosion.

The erosion controlling factor is vegetation retention, which determines whether vegetation is present on cropland or grasslands throughout time to protect soil from erosion. The following equation is used to calculate this factor (24):

$$V_m = \exp\left[10 \cdot \left(1 - BS_f\right) \cdot Fcover_m\right] \quad (6)$$

where $V_m$ is the monthly vegetation retention factor composite, $BS_f$ is the bare soil frequency where $1 - BS_f$ indicates the vegetation cover frequency, $Fcover_m$ is the monthly green vegetation cover fraction composite, and $m$ is the computation month. The annual vegetation retention factor ($n$) is calculated by adding the monthly vegetation retention factors over the monitoring period.

$$V = \frac{\sum_{i=1}^{i=n} V_{m,i}}{n} \quad (7)$$

where $V$ is the annual vegetation retention. Using Sentinel-2 imagery data, we estimated $Fcover_m$ employing the Google Earth Engine (GEE) platform and $BS_f$ using the Geospatial Soil Sensing System (GEOS3) algorithm [estimated by Normalized Vegetation index (NDVI) and Normalized Burn ratio 2 (NBR2)].

#### 4.1.2. Landscape effect (L)

The landscape factor is associated with the practice management factor as expressed in the RUSLE equation. This factor incorporates the effect of landscape features in controlling erosion (*e.g.*, contouring) and interrupting the rainfall-runoff effect (denoted by ICP and WCP prefixes, Table S4, Fig. S4) in National Resource Inventory (NRI) water erosion estimations (27). Karydas and Panagos, (24) proposed a formula to estimate this factor using a 3×3 Sobel filter (an edge-detection filter) applied to the Near-Infrared (NIR) band of Sentinel-2 as follows (24):

$$L = 1 + \sqrt{\frac{s_f}{DN_{max}}} \quad (8)$$

where, $L$ is the landscape effect (dimensionless), $s_f$ is the Sobel filter value (dimensionless), and $DN_{max}$ is the maximum potential reflectance value (which is 10000 in Sentinel-2 data on GEE).

While this methodology can be applied to croplands, on a national level, additional factors for non-cropland land types (*e.g.*, forested lands, urban areas, and grass/pasturelands) are necessary. These were gathered from the literature (2, 3, 28, 29) — see Table S2.

## 4.2. Present Sustainability Assessment

We calculated the sustainability factor using Sentinel-2 imageries using various algorithms in the GEE platform. To mask non-cropland areas, we used various thresholds to mask clouds, shadows, and bare lands (*e.g.*, urban areas or desert lands), and water. First, we used a 40% probability of cloud to mask the image from the calculation, 60% probability of cloud in Sentinel-2 metadata, a 0.15 (~on band scale) NIR to determine whether shadows of clouds should be considered, and 15 pixels distance and/or 50-m buffer from cloud masked pixels. Therefore, these thresholds mask clouds and shadows from Sentinel-2 images. Second, a GEOS3 algorithm was used to mask bare soils (>95% of occurrence) in each image of Sentinel-2 masked in the previous step. Third, urban areas were masked using the Facebook population layer and Global Human Settlement Layers (GHSL). Fourth, we masked out water from the sustainability factor estimation using the max extent of water from the JRC Global Surface Water Mapping Layers database. Finally, the masked collection of Sentinel-2 images was prepared to estimate vegetation retention and landscape effect. We used masked Sentinel-2 images to estimate FCOVER using (Sentinelhub custom-scripts repository). Then, we generated a list of time series to prepare a composite of Sentinel-2 images in a monthly period and used a harmonic regression (Fourier transform) to prepare an estimation of FCOVER in annual resolution. To estimate the bare soil frequency, we used bare soil images (~calculated by the GEOS3 algorithm in previous steps) and estimated the frequency (<95% of occurrence). Then, we used the median composites of Sentinel-2 images to estimate the landscape effect using a 3×3 Sobel filter (an edge-detection filter) applied to the NIR band of Sentinel-2. Finally, the composites of FCOVER factor were integrated during the period of estimation (in 2020) to calculate vegetation retention. Accordingly, the sustainability factor was estimated.

The proposed method could estimate the sustainability factor at a 10-m resolution. However, cloud coverage, shadow masking, and Sentinel-2 revisiting time (~5-10 days) made some images unavailable/unsuitable in some regions to estimate the sustainability factor. This led us to use covariates to estimate the sustainability factor across the contiguous United States. Therefore, we developed a Random Forest (RF) Machine Learning (ML) model to interpolate the sustainability factor. To do so, we used the National Agriculture Imagery Program (NAIP, 0.6-m) (30), Net Primary Production (NPP, 30-m) and Gross Primary Production (GPP, 30-m) (31) to resample the sustainability factor to a 30-m resolution across various croplands identified by the CropScape (~30-m) database of the United States Department of Agriculture (USDA) (32) and total croplands extent using ESA World Cover (~10-m). Then, ESA World Cover was used to assign literature values of non-croplands (see Table S2). We trained the RF model using 65000 points across the contiguous United States using 200 trees, with the number of leaves set to unlimited, to estimate the sustainability factor in an iterative manner. Finally, the sustainability factor was mapped across the contiguous United States (Fig. S8A).

### 4.3. Future Sustainability Projection

Future land use and land cover (LULC) projections proposed by Land Use Harmonization (LUH2) and PFT-based cellular automata (PFT-CA) based on CMIP6 scenarios were used to project sustainability factors. However, LUH2 (~5.5 km) and PFT-CA (~1-km) coarse resolutions restricted the high-resolution projection of the sustainability factor. Therefore, we used a RF model similar to that proposed for baseline assessment except using LULC models instead of NAIP for projections. To downscale future sustainability factors, NPP and GPP data from 2015 were used along with future scenarios of LUH2 and PFT-CA LULC models. This method was applied to cropland areas, while non-cropland area sustainability factors were assigned from the literature (see Table S2). Therefore, sustainability factors were projected across the contiguous United States using three alternative (2.6, 4.5, and 8.5) Shared Socioeconomic Pathway and Representative Concentration Pathway (SSP-RCP) scenarios of CMIP6 (Figs. S8B-D).

# 5. Terrain

While several methods exist for calculating the influence of terrain (~slope length and steepness factor of RUSLE), these are limited by the complexities of landscape (*e.g.*, steep slope angles or short slope length). This study used the unit-contributing area concept to calculate the slope length and steepness factor (33). This method subdivides the slope into sub-segments, modifying slope steepness into more uniform shapes. It is applied to various regions, from steep mountain slopes to gentle slopes on the plains. We used the unit-contributing area concept and the original equation proposed by Desmet and Govers, 1997 to derive the Terrain influence factor as follows (33):

$$\begin{cases} S = 10.8 \times \sin(\theta) + 0.03, & slope < 9\% \\ S = 16.8 \times \sin(\theta) - 0.05, & slope \geq 9\% \end{cases}$$

$$L_{i,j} = \frac{(A_{i,j} + D^2)^{m+1} - A_{i,j-in}{}^{m+1}}{D^{m+2} \times x_{i,j}^{m} \times 22.13^{m}}$$

$$m = \frac{\beta}{\beta + 1} \tag{9}$$

$$\beta = \frac{\frac{\sin(\theta)}{0.0896}}{0.56 + 3 \times [\sin(\theta)]^{0.8}}$$

where $S$ is slope steepness, $\theta$ is the gradient of the slope (°), $L_{i,j}$ is slope length (m), $A_{i,j}$ is upstream contributing area at the inlet of the grid cell $(i,j)$ (m$^2$), $D$ is grid cell size $(m)$, $x_{i,j} = \sin(a_{i,j}) + \cos(a_{i,j})$, $a_{i,j}$ is the aspect direction of the grid cell $(i,j)$, and $\beta$ is the ratio of rill to inter-rill erosion. To estimate the LS-factor, we used the GEE platform using the contributing area drawn from the MERIT database (~90-m) and the National Elevation Dataset (NED, ~10-m) because the coarse resolution would lead to overestimation of the topography effect. The MERIT contributing area was resampled to a 10-m resolution and terrain influence was calculated at a 10-m resolution, using a combination of MERIT and NED databases (Fig. S9).

# SI Results and Discussion

# 1. Evaluation

## 1.1. Rainfall Erosivity

We validated and compared our rainfall erosivity estimation with previously available datasets (Figs. S3, A-D). Our results show the GloREDa dataset (Fig S3B) had the lowest accuracy for rainfall erosivity estimation, and that this was mainly related to the number of gauges used (~92), a number not comparable with the 3221 stations used in the present study. We found the rainfall erosivity of RUSLEv2 calculated by Renard et al., 1997, (10; Fig. S3C) between 1975 and 1995 to be overestimated in terms of the short estimation period compared with the present study (1971-2013) and its time resolution of 30-minutes. However, the linear trends between RUSLEv2 and our calculated rainfall erosivity were comparable. We also used an Empirical Bayesian Kriging (EBK) to interpolate rainfall erosivity (Fig. S3D). A simple interpolation model achieved better accuracy than a DL model; however, the EBK model underestimated high values of rainfall erosivity. The assessment of DL and EBK models indicated that rainfall erosivity interpolation was comparable with calculated values, while RUSLEv2 and GloREDa showed some deviations with calculated values— mostly overestimations.

Climate factors play a crucial role in environmental issues. Rainfall erosivity modeling is a complex process, affected by the duration of assessment and observation time interval, *e.g.*, 15 minutes interval. Although previous studies have tried to model this factor at a regional or global scale, their simplifications and the low number of stations used to supply data usually led to uncertain estimations. The accuracy of the proposed method in interpolating rainfall erosivity by a non-stationary DL model integrated with a GPR ML kernel in estimating rainfall erosivity was promising [$R^2 = 0.84$, total RMSE = 840 MJ $mm^{-1}$ $ha^{-1}$ $yr^{-1}$].

Overall, GloREDa overestimated rainfall erosivity by about 8% compared to our study. Comparing the maps between GloREDa and our estimations (Fig S3A,B) shows different overall patterns, especially in the western deserts where monsoon rainfalls are common. The few stations used by GloREDa in these areas led to substantial differences in real rainfall erosivity compared with our estimations.

Another influential factor affecting rainfall erosivity is elevation. As the AdaptWest dataset was characterized by SRTM DEM (1-km resolution), estimations based on this dataset considered precipitation, humidity, snow, evaporation, and geospatial location and reflected the elevation effects on the DL model, integrated with GPR interpolation. The GloREDa output showed certain inconsistencies with our estimations in the western mountains, mostly in terms of elevation dependence structure. Results indicated that with rising elevation, rainfall erosivity would decrease in a non-linear pattern. We show the rainfall erosivity and associated parameters from rainfall erosivity estimation between 1971 and 2013 (Fig. S10).

Additionally, we compared our results to those of a similar study encompassing the contiguous United States (34), and found that while our results were comparable, their study overestimated rainfall erosivity even when the time resolution (15-min) was decreased. Yin et al. (36), showed that by increasing the time resolution, the rainfall erosivity decreased.

## 1.2. Sustainability Evaluation

The sustainability factor (*e.g.*, a combination of cover and practice management in RUSLE) is a complex factor to estimate. Previous studies mostly proposed a static value for each crop type (Table S2), comparable with the proposed method. However, these values may change from point to point and also be over- or under-estimated when they are defined for each crop statically in terms of agricultural density, intensity, bareness, extension, and management practices used for each crop in different periods and geospatial locations around the world and CONUS (3). Hence, for each pixel, the proposed method predicted a sustainability factor and considered the management practice (*e.g.*, tillage, crop rotation, bedding, etc.). This resulted in a substantial improvement in estimating sustainability factors compared with static values assigned on the basis of land use and land cover maps with a margin of errors in classification. Therefore, corn crops were assigned different values in various states, mainly related to the type of practices used to control soil erosion (Table S4).

## 1.3. Soil Erosion Evaluation

Soil erosion estimation becomes uncertain when modeling is conducted at coarse resolutions and also when management practices are ignored (35). Soil erosion happens through small-scale processes such as sheet and rill erosion. Estimations at coarser resolutions have been shown to be too bulky for national management scales, being so large that mesoscale processes are ignored. Therefore, a 30-m resolution was used in this study, so that its outputs would prove useful to policymakers operating with actual data at field management scales. Previous studies suggested that validation of regional models. such as FSSLM, with *in situ* measurements is challenging due to the lack of long-term and comparable spatiotemporal resolution. However, we compared the FSSLM model with various *in situ* measurements and models to help policymakers support the plausibility of the model in the field, regional, and continental scales.

### 1.3.1. FSSLM cross-comparison with the National Resource Inventory (NRI)

The first comparison of FSSLM results, at national and states scales with NRI results occurred in 2017. It showed the plausibility of the FSSLM model by comparison with *in situ* measurements. On a national scale, cropland soil erosion in 2017 (sheet and rill erosion), was estimated by the NRI to be 0.979 Pg $yr^{-1}$, which was comparable to FSSLM results (0.784 Pg $yr^{-1}$). These deviations between FSSLM and NRI estimations (Fig S13A) were rooted in various factors employed in the NRI estimation: (*i*) NRI employed the RUSLEv2 rainfall erosivity factor, which was significantly greater than our estimations (see Rainfall Erosivity section, 38), (*ii*) NRI data were gathered from a set of 200000 sampling sites, which was not comparable to our study's resolution with 1.84 billion raster points at a resolution of 30-m — and capable of rising to 4.61 trillion raster points if NAIP images are used for soil erosion estimation (could be performed with the proposed methodology). Therefore, for example, in Tennessee and Kentucky, soil erosion estimates are greater than NRI estimations, whereas in Iowa and Illinois, the contrary is true (Fig. S13A). However, results in states with lower soil erosion values are comparable because of the low extent of croplands.

### 1.3.2. FSSLM cross-comparison with the Daily Erosion Project (DEP)

We compared our results with the Daily Erosion Project (DEP) (39; Fig S13B). The DEP estimates daily soil erosion in the HUC-12 (*e.g.*, Iowa, and parts of Kansas, Missouri, Nebraska, and Minnesota) using the Water Erosion Prediction Project (WEPP) model that estimates soil loss on hillslopes employing Next-Generation Weather RADAR (NEXRAD) precipitation data. The DEP draws its cover and practice management (sustainability) factor from confidential NRI databases.

The comparison of the FSSLM and DEP models between 2007 and 2020 in the HUC-12 shows reasonable agreement (Fig. S13B) with a 1.76 Mg $ha^{-1}$ $yr^{-1}$ mean absolute error. The differences were rooted in the time frames of weather data in the DEP model between 2007 and 2020 and in the present study between 1971 and 2013. Differences in agricultural land areas also made this not a one to one comparison. Although NRI estimates show higher soil erosion rates and the DEP model lower values, these are more reasonable in terms of previous limitations cited for NRI estimations.

# 2. Model limitations

Based on the G2 erosion model that estimates soil erosion, the FSSLM model includes sheet and rill erosion processes, but excludes gully and alluvial erosion processes. However, it remains the best available method to estimate soil erosion and, when combined with sediment yield equations (models), can estimate sediment transport (3, 29, 38). Although we used the finest resolution dataset available for the CONUS, uncertainties in land cover maps, station rainfall measurements, soil database accuracy, and elevation dataset precision led to inaccuracies not associated with our methodology but related to the data itself. The G2 erosion model inherits limitations of the RUSLE equation in terms of sheet and rill erosion, including: (*i*) no modeling of degradation; (*ii*) no modeling of alluvial erosion; (*iii*) ignoring gully erosion; (*iv*) the rainfall erosivity factor is defined and measured based only on the eastern regions of the United States; (*v*) the soil erodibility is adjusted to the eastern United States; (*vi*) steep slopes have some extreme LS factor values that cannot be resolved by any method, although some equations try to relax these factors (39); and (*vii*) interdependency of factors is ignored in the original RUSLE (40), but are considered in the modified version 2 (41). Notably, the G2 erosion model is very complex, implying a method designed for continental scales.

The limitations of the G2 erosion model have encouraged the authors to use better datasets, including the high-resolution land cover dataset (CropScape, 30-m), satellite imageries and products, the high-resolution climate dataset (AdaptWest 1-km), the soil properties dataset (POLARS 30-m and SoilGrids 250-m), and the elevation dataset (USGS NED 10-m and MERIT 90-m), to estimate soil erosion. Our estimation was prone to various uncertainties and errors that better datasets could solve, and was not related to methodological procedures or processes. We ignored the effects on soil erosion of wildfire, greenhouse emissions that change soil erosion side effects on soil erodibility, and soil microorganisms. We included simplified seasonal vegetation, double vegetation, forestland transition, and urban area erodibility. These factors had substantial effects on the soil erosion model. However, reliable datasets and the complexity of estimation for continental scales limited our ability to provide a more realistic dataset. Soil erosion modeling based on improved datasets, which will be available in the future, is a goal for subsequent research.

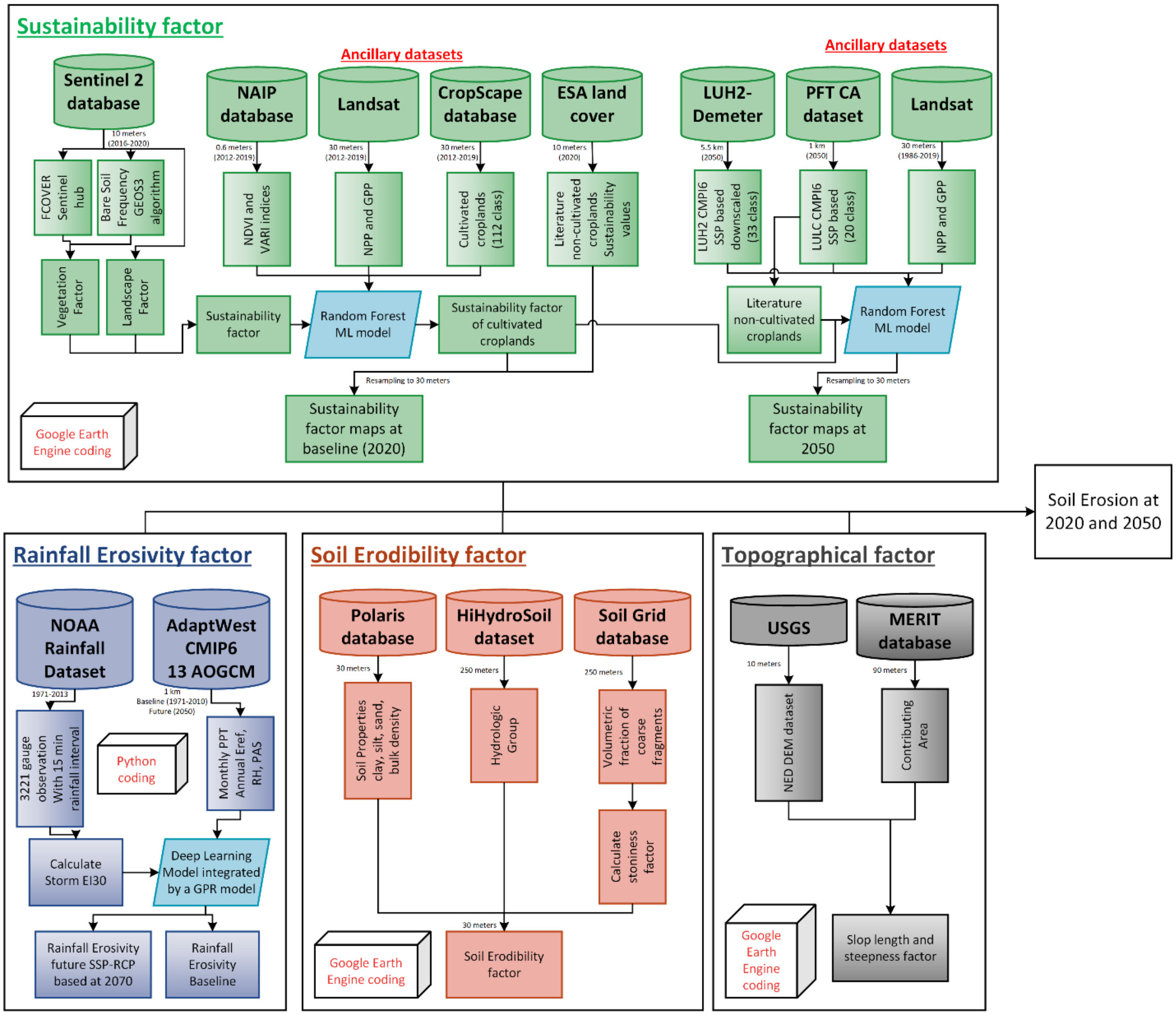

**Figure S1** Flowchart of the FSSLM.

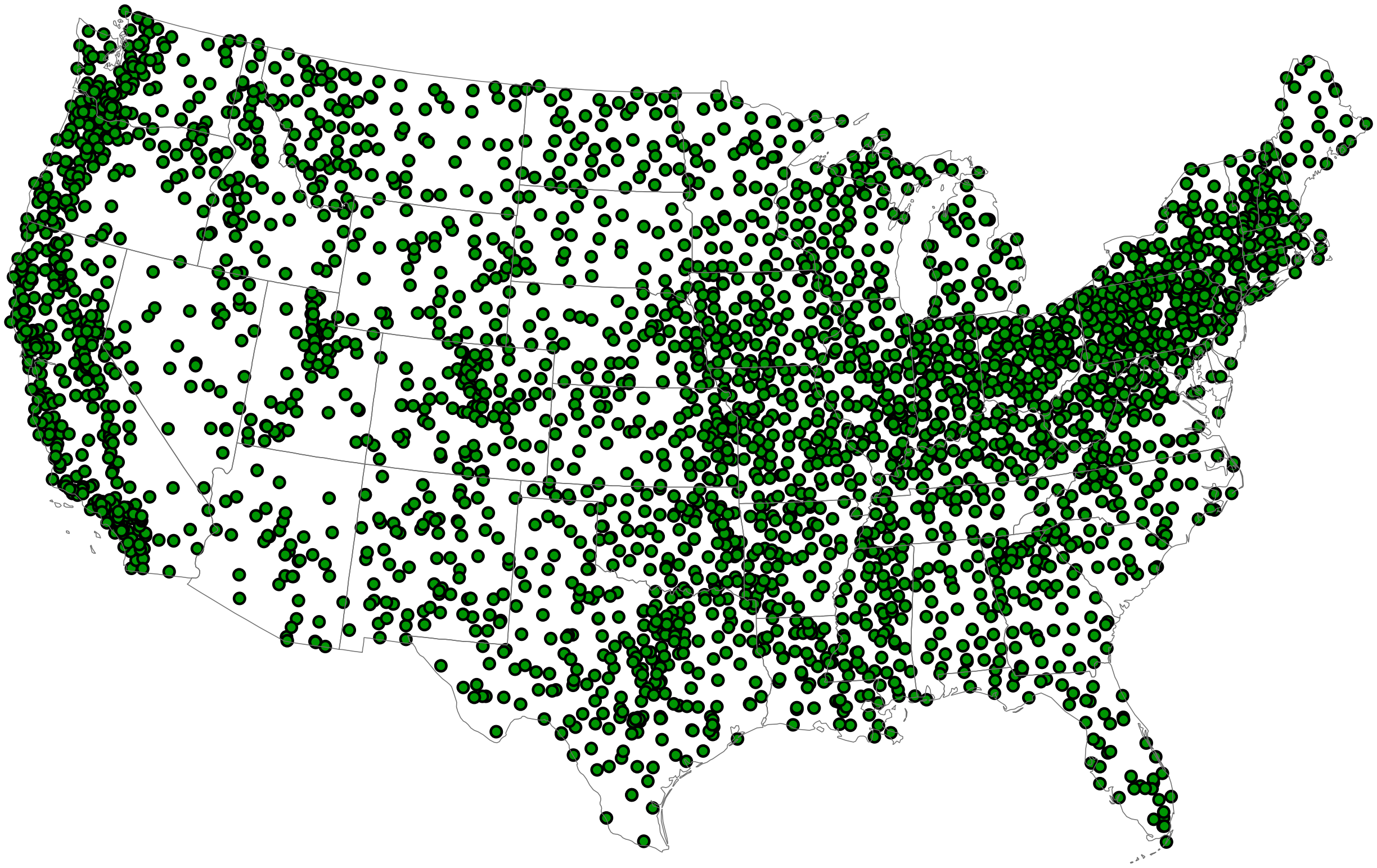

**Figure S2** Active NOAA precipitation stations (*n* = 3221) with 15-minute interval data between 1971 and 2013.

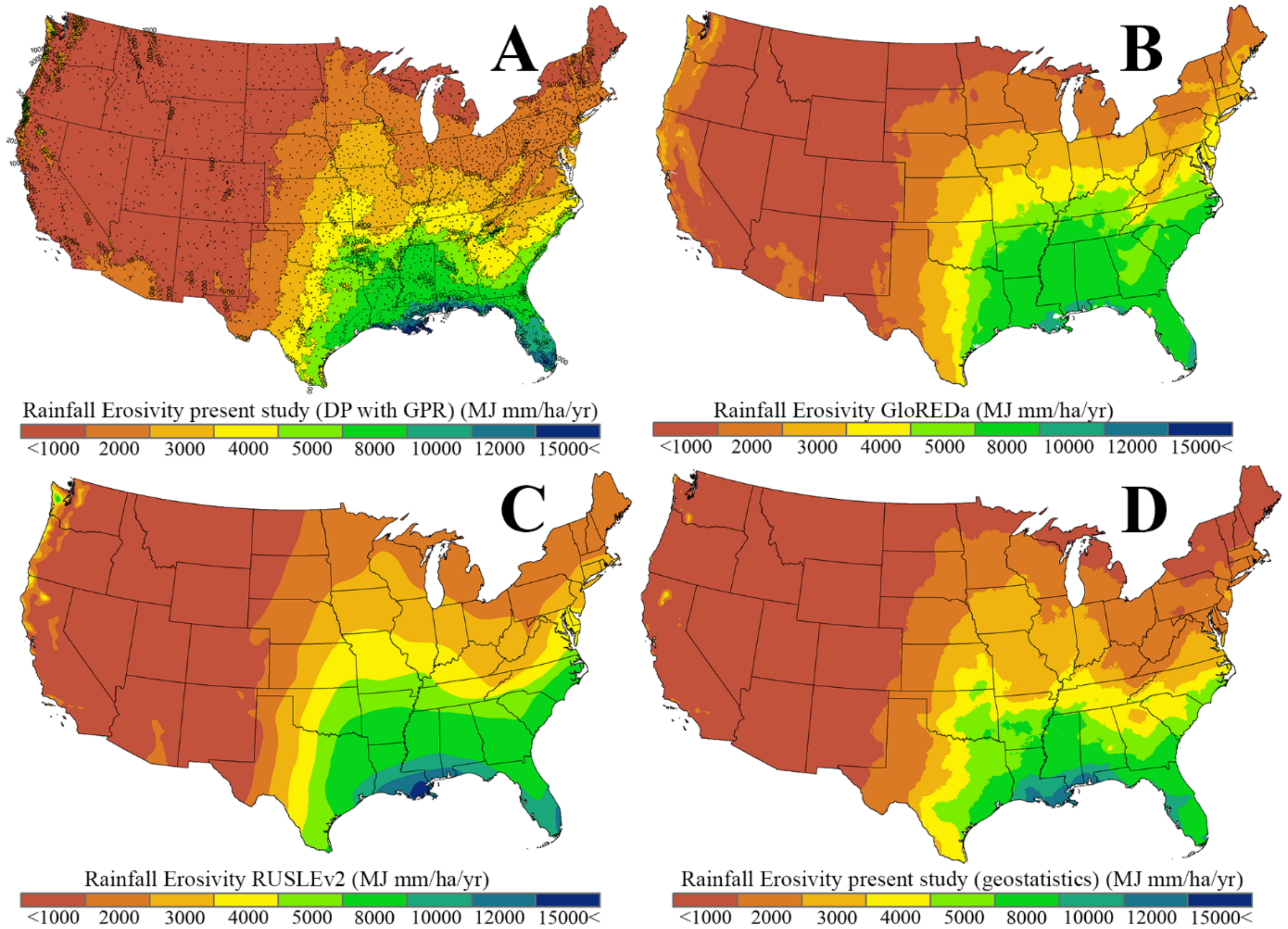

**Figure S3** Rainfall erosivity maps of the United States (MJ mm $ha^{-1}$ $yr^{-1}$): (A) present study, (B) GloREDa database, (C) RUSLEv2, and (D) Empirical Bayesian Kriging (EBK).

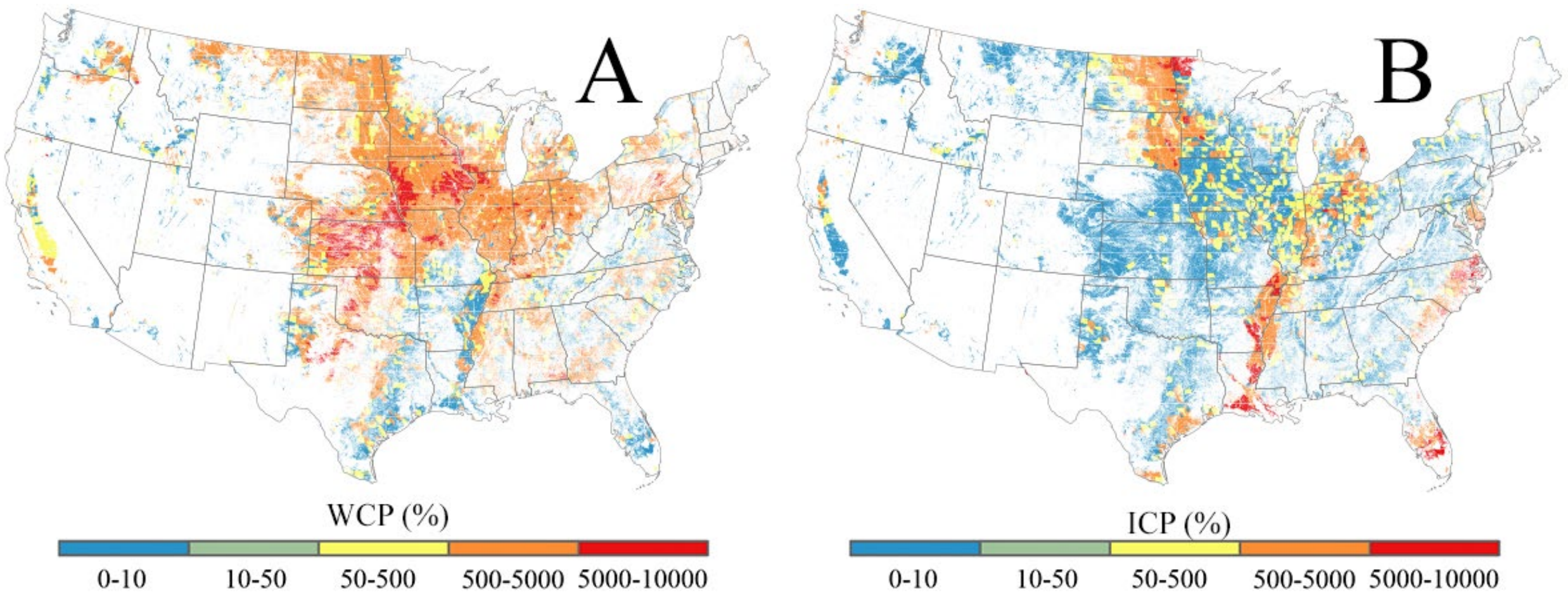

**Figure S4** Maps of management practices (percentage of agricultural area), (A) WCP, (B) ICP.

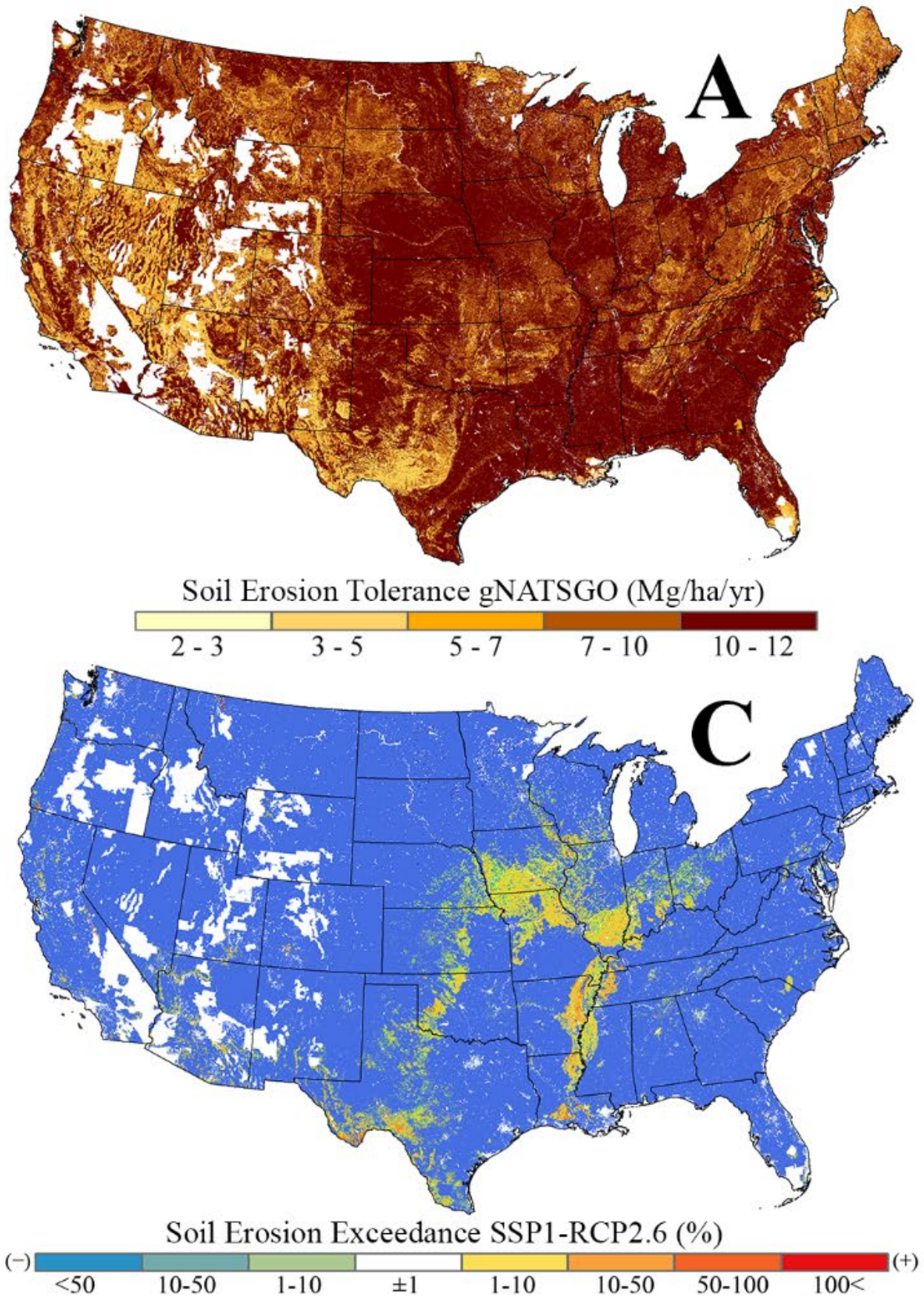

**Figure S5** (A) Soil erosion tolerance (Mg $ha^{-1}$ $yr^{-1}$), (B) soil erosion exceedance at baseline (%), (C) soil erosion exceedance under SSP1-RCP2.6 (%), and (D) soil erosion exceedance under SSP5-RCP8.5 (%), from soil erosion tolerance.

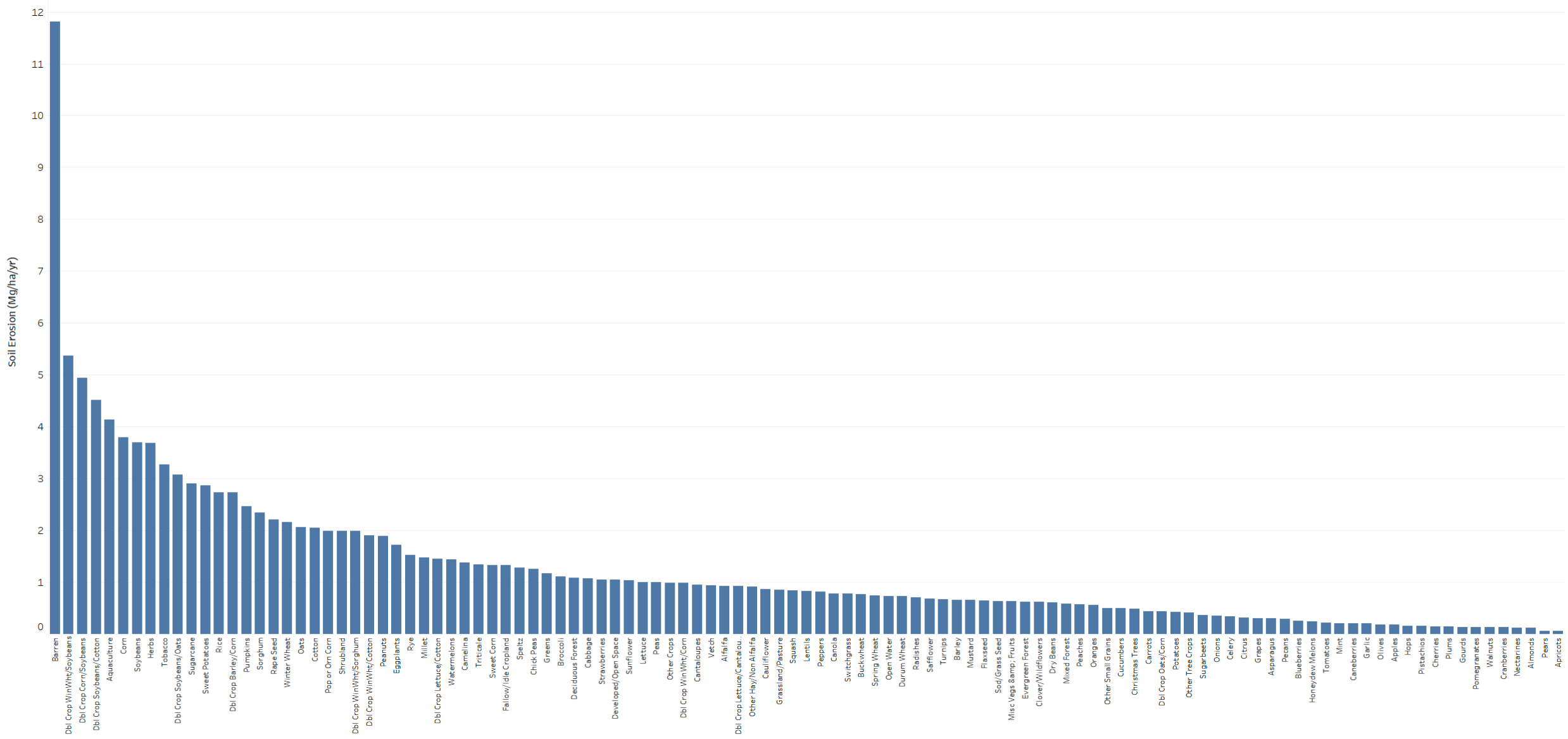

**Figure S6** Soil erosion in CropScape database LULC classes.

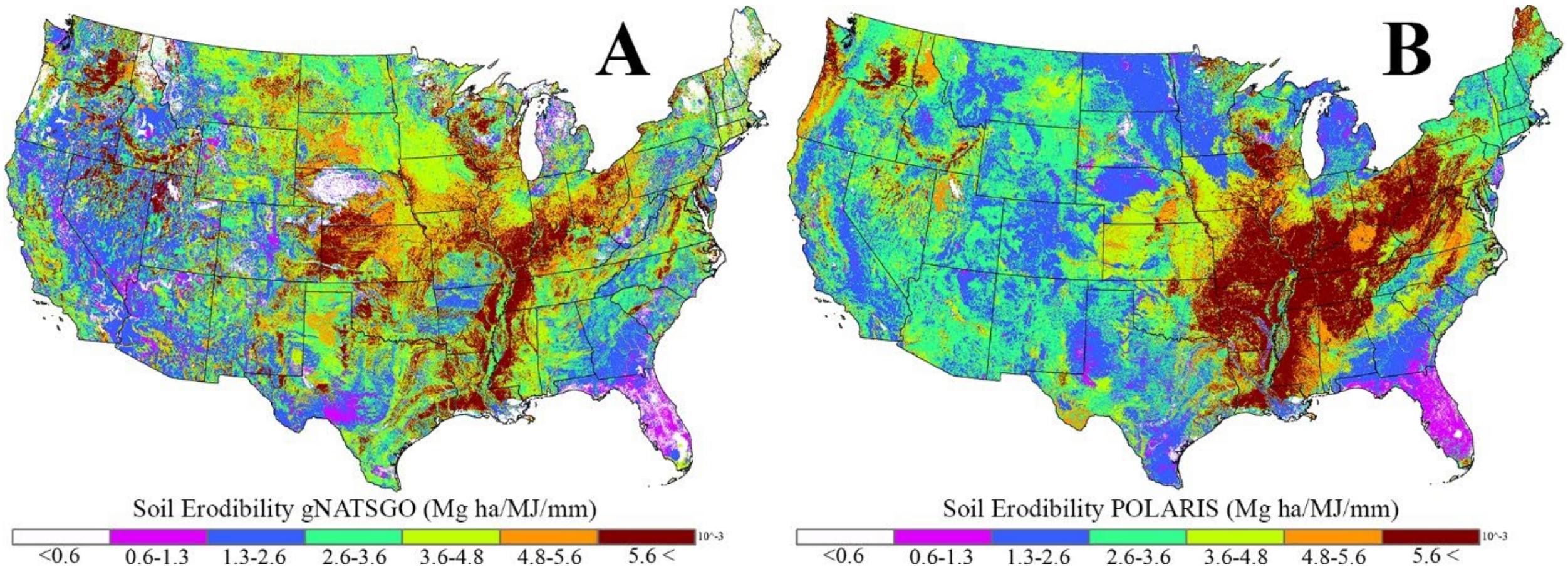

**Figure S7** Soil Erodibility (Mg ha $MJ^{-1}$ $mm^{-1}$) maps of the United States, (A) gNATSGO, and (B) POLARIS.

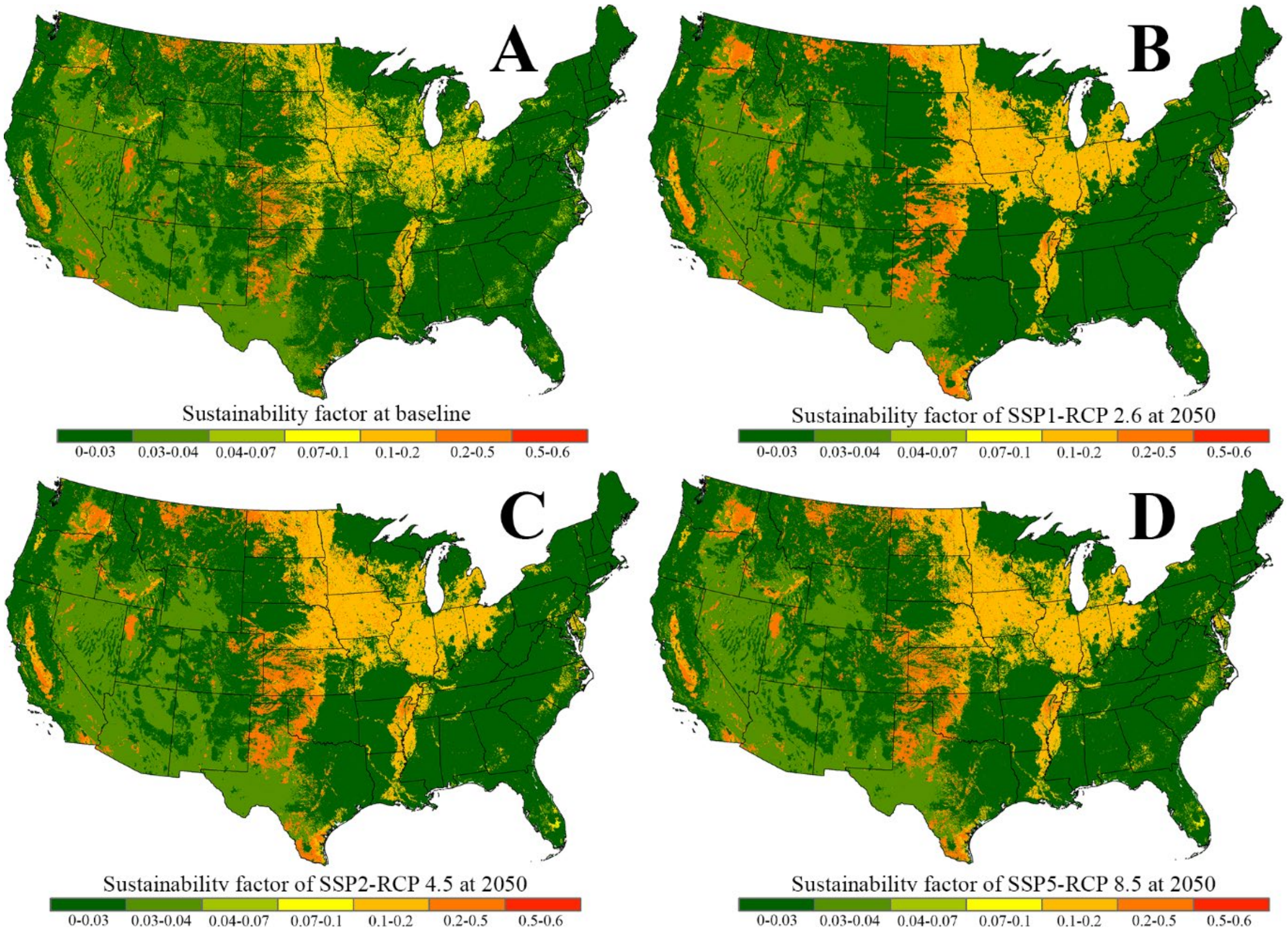

**Figure S8** Sustainability factor maps: (A) Baseline, (B) SSP1-RCP 2.6, (C) SSP2-RCP 4.5, and (D) SSP5-RCP 8.5 scenario in 2050. The red color bar shows the highest cover management factor, which means more likely to erode, while the green color shows the lowest cover management factor and less likely to erode.

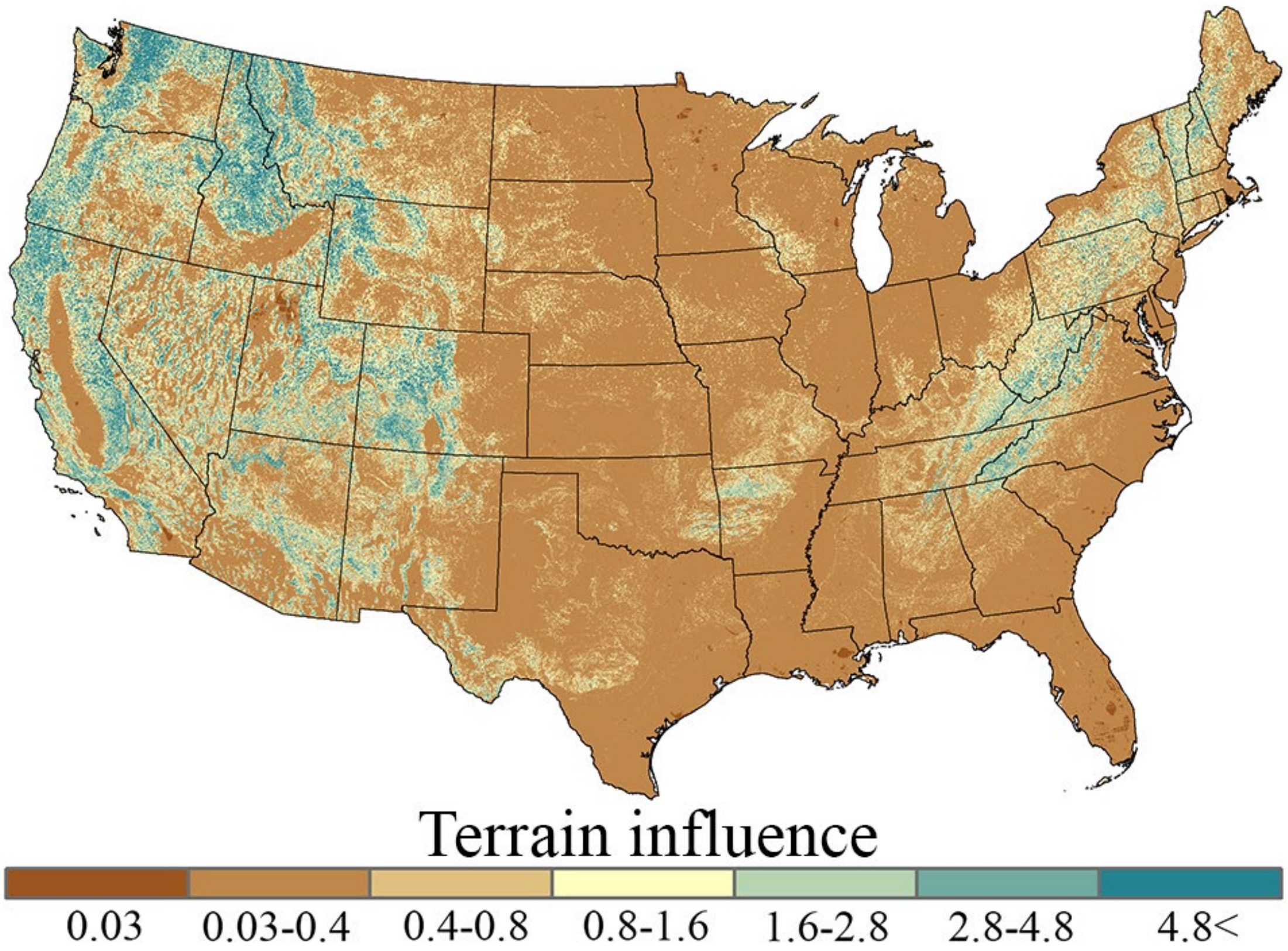

**Figure S9** Terrain influence.

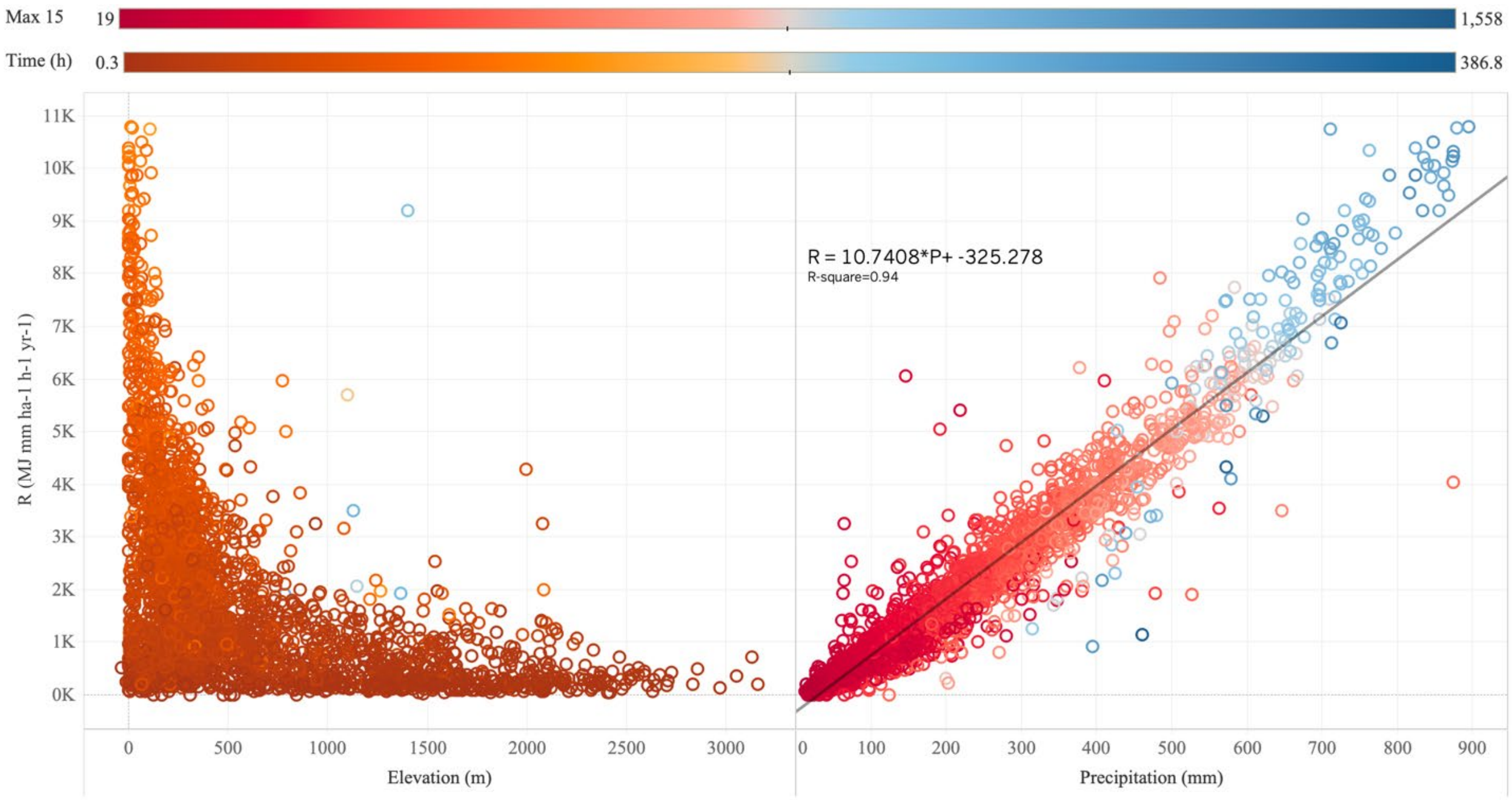

**Figure S10** Rainfall Erosivity (R-factor) versus (left) elevation, (right) total precipitation. The upper color bar shows the maximum intensity in 15 minutes, and the lower color bar shows the time of events in hours.

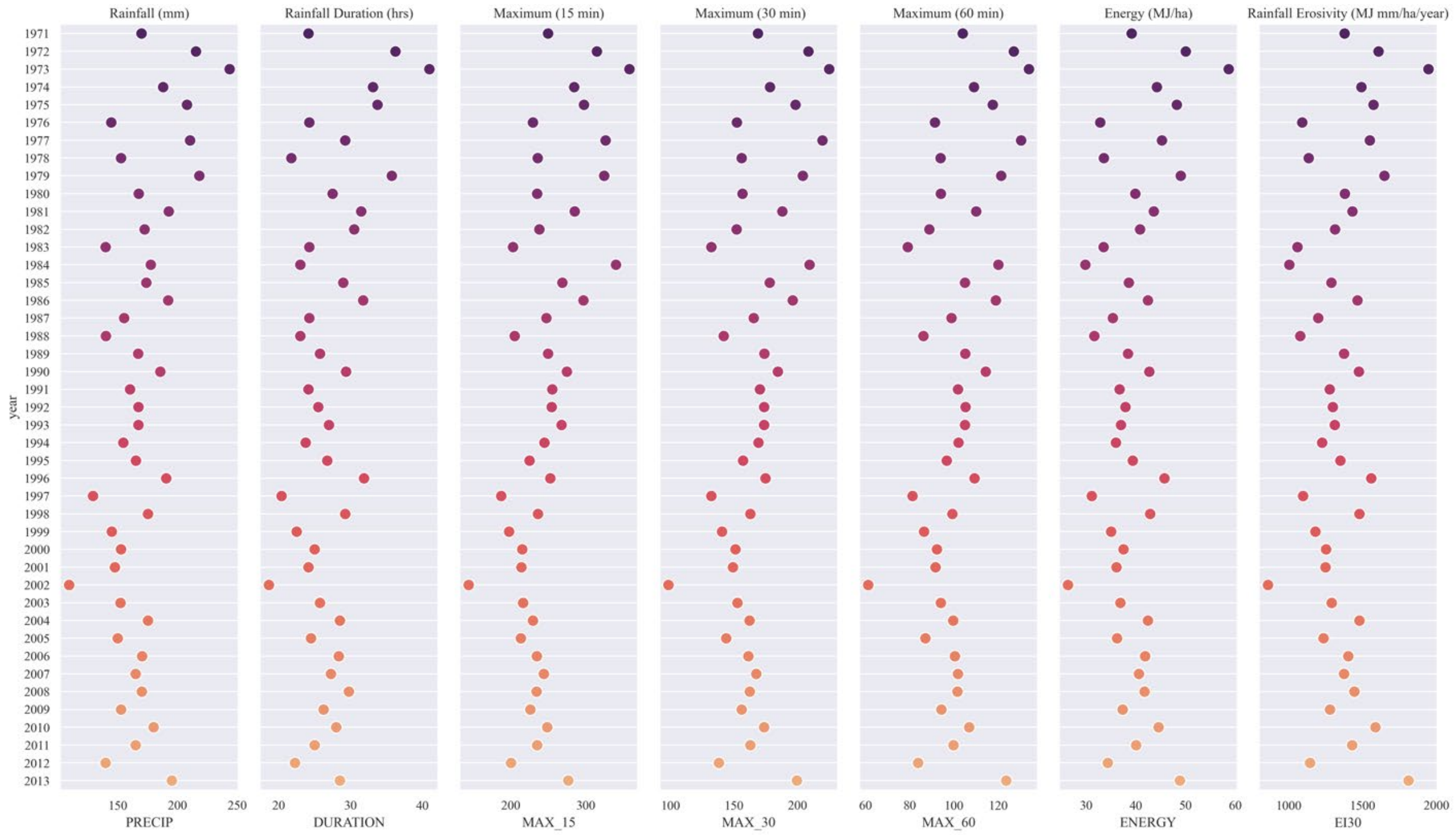

**Figure S11** Median Rainfall (mm), Rainfall Duration (hrs), Maximum (15 min), Maximum (30 min), Maximum (60 min), Energy (MJ $ha^{-1}$), and Rainfall Erosivity (MJ mm $ha^{-1}$ $yr^{-1}$) in each year between 1971 and 2013.

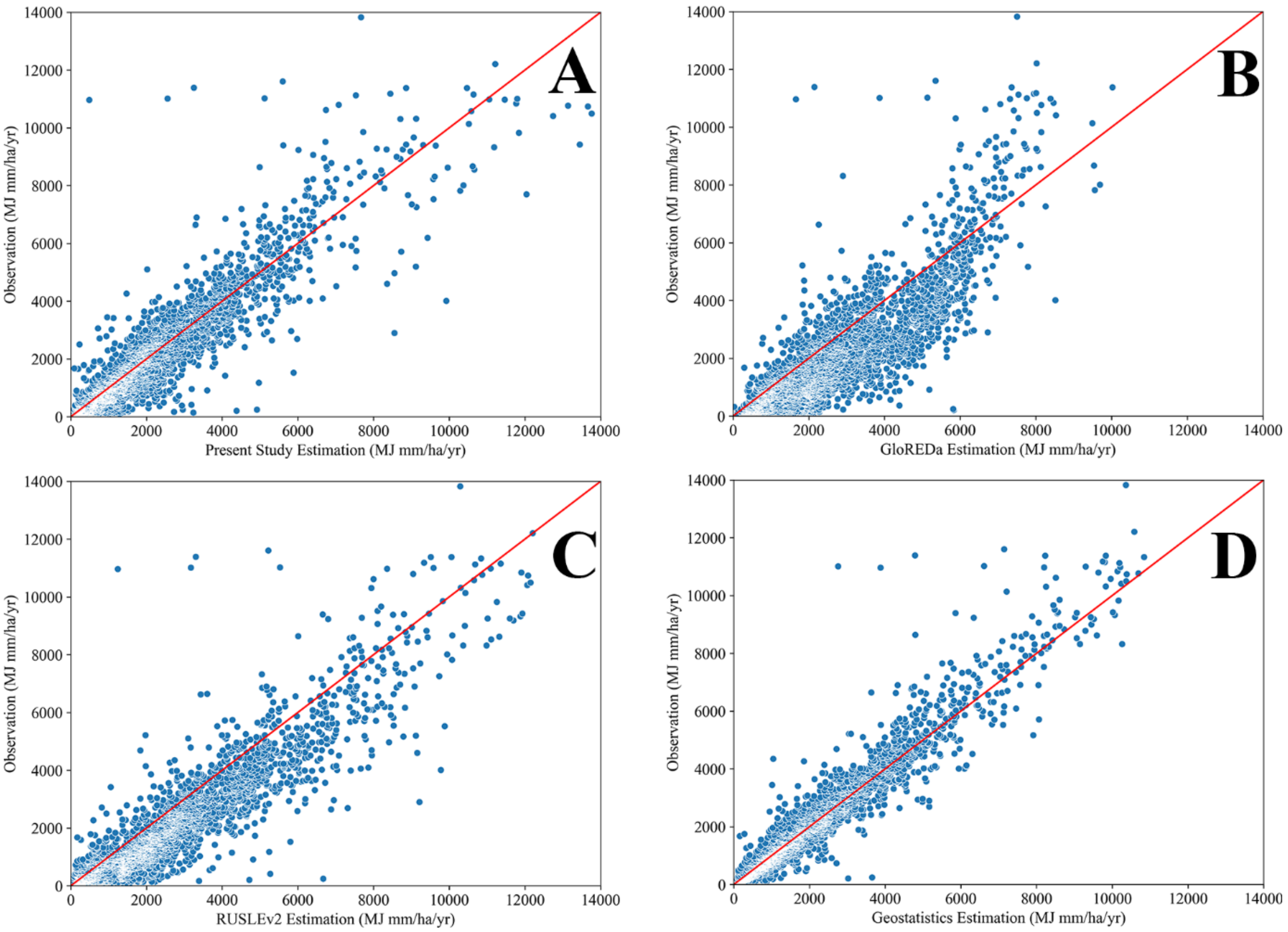

**Figure S12** Rainfall erosivity linear regression between calculated rainfall erosivity of 3221 active NOAA 15-min interval precipitation gauges and (A) present study, (B) GloREDa database, (C) RUSLEv2, and (D) Empirical Bayesian Kriging (EBK).

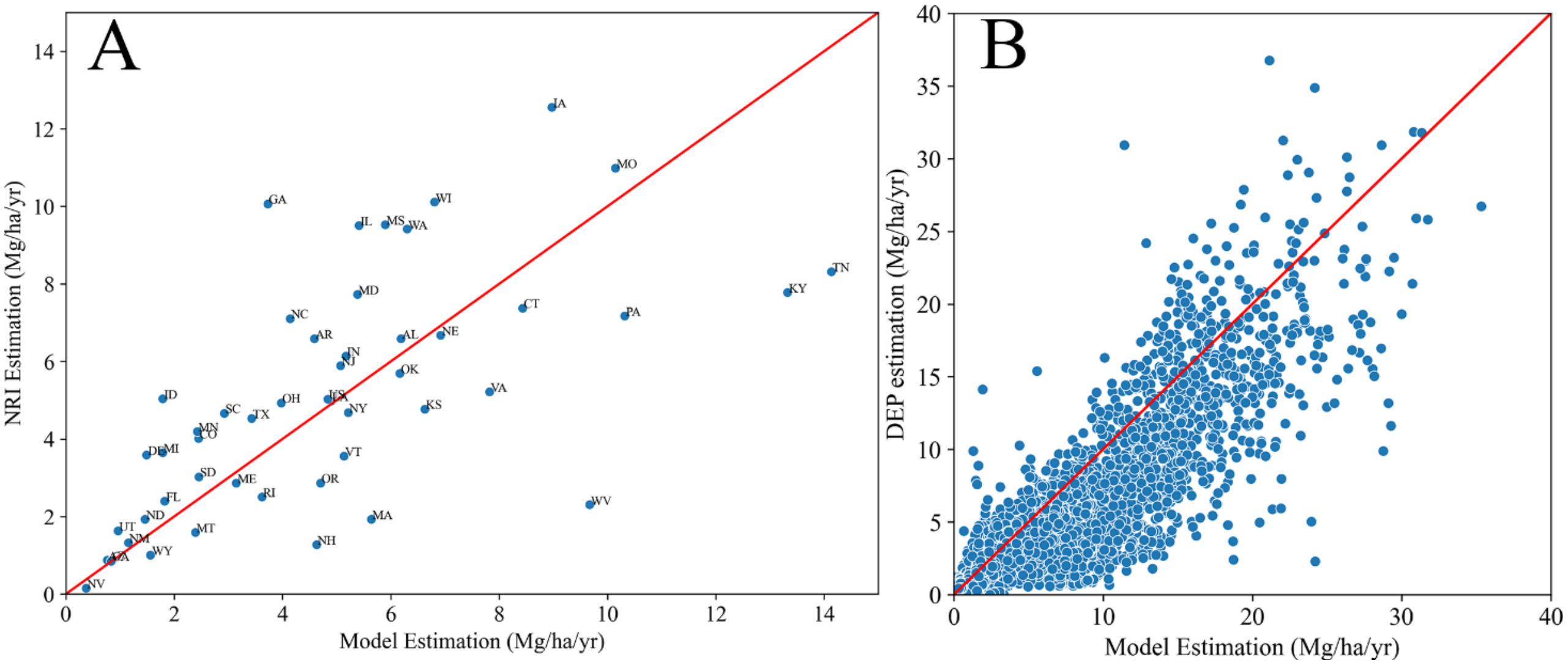

**Figure S13** Soil erosion comparison in (A) between NRI estimates and our study by state, (B) bewteen DEP and our study by Hydological Units in Iowa, and parts of Kansas, Missouri, Nebraska, and Minnesota.

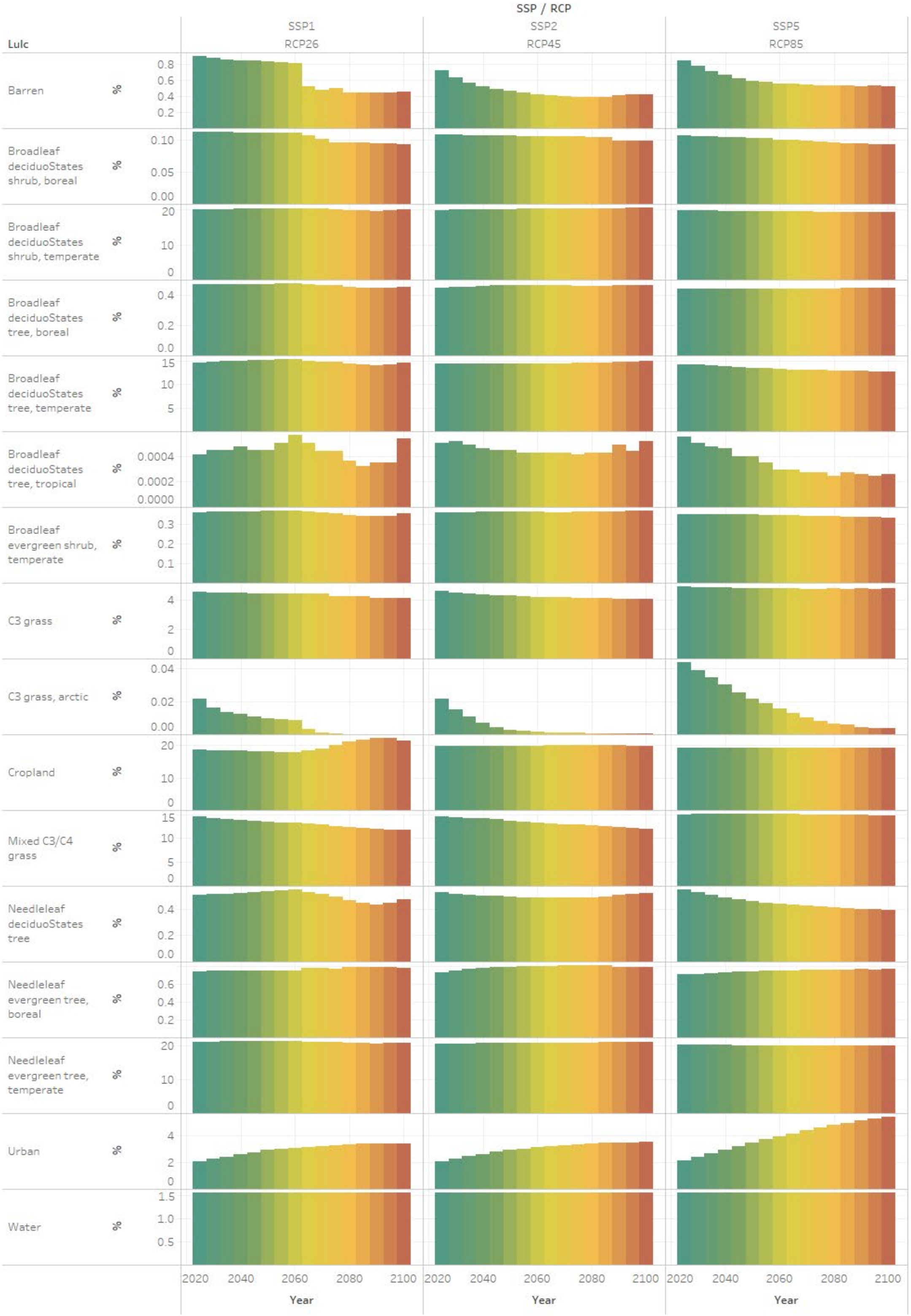

**Figure S14** Land use percentage of different land cover classes in different decades and for four scenarios plus a historical baseline.

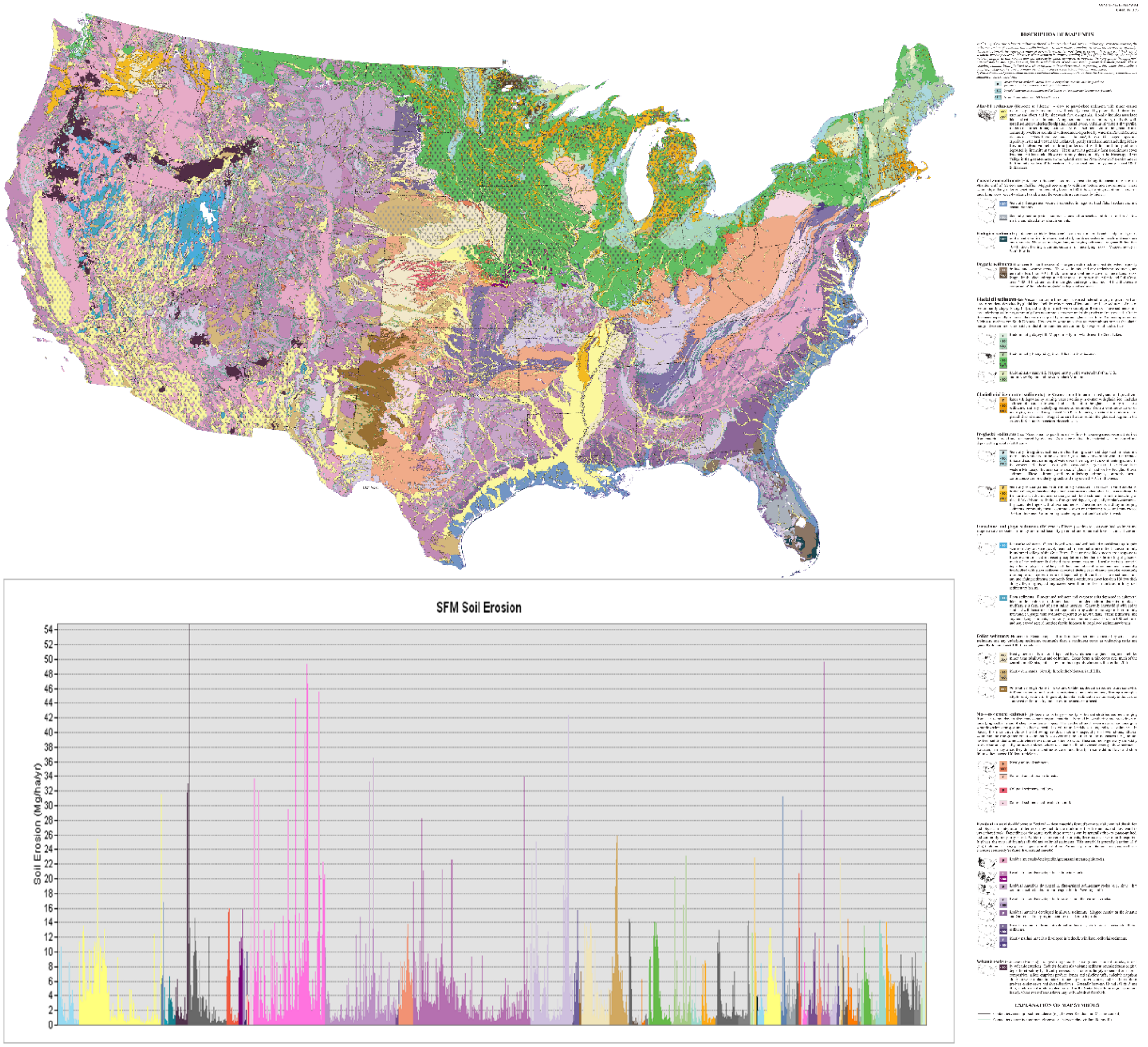

**Figure S15** Soil erosion in sedimentology classes across the CONUS (adapted from (42)).

**Table S1.** Average cover management factor of non-croplands (for NLCD,MODIS PFT, and ESA types) (2, 3, 28, 29).

| NLCD | MODIS | ESA | C factor |
|---|---|---|---|
| Open water | Water | Open water | 0 |
| Perennial ice/snow | Permanent snow and ice | Snow and ice | 0 |
| Developed, open space | Urban | Built-up | 0.03 |
| Developed, low intensity | | | 0.03 |
| Developed, medium intensity | | | 0.03 |
| Developed, high intensity (Mining) | | | 0.03 |
| Deciduous forest, Evergreen forest, Mixed forest | Broadleaf evergreen tree, tropical<br>Broadleaf evergreen tree, temperate<br>Broadleaf deciduous tree, tropical<br>Broadleaf deciduous tree, temperate<br>Broadleaf deciduous tree, boreal<br>Needleleaf evergreen tree, temperate<br>Needleleaf evergreen tree, boreal<br>Needleleaf deciduous tree | Trees<br>Mangroves<br>Moss and lichen | 0.002 |
| Shrub/scrub | Broadleaf evergreen shrub, temperate<br>Broadleaf deciduous shrub, temperate<br>Broadleaf deciduous shrub, boreal | Shrubland | 0.03 |
| Barren Land | Barren | Barren / sparse vegetation | 0.29 |
| Grassland/herbaceous/Pasture | C3 grass, arctic<br>C3 grass<br>C4 grass<br>Mixed C3/C4 grass | Grassland | 0.07 |
| Woody wetlands | - | - | 0.001 |
| Emergent herbaceous wetlands | - | Herbaceous wetland | 0.001 |

**Table S2.** Sustainability factor of croplands (summary) compared with literature (2, 5, 43–48).

| Cropland | Sustainability | Literature |
|---|---|---|
| Alfalfa | 0.21 | 0.1-0.15 |
| Almonds | 0.21 | 0.25 |
| Apples | 0.24 | - |
| Aquaculture | 0.15 | - |
| Barley | 0.17 | 0.21 |
| Blueberries | 0.26 | 0.15-0.2 |
| Broccoli | 0.13 | 0.25 |
| Canola | 0.14 | - |
| Cantaloupes | 0.32 | 0.25 |
| Cherries | 0.28 | - |
| Chick Peas | 0.41 | 0.32 |
| Citrus | 0.38 | - |
| Clover/Wildflowers | 0.19 | 0.10-0.15 |
| Corn | 0.13 | 0.38 |
| Cotton | 0.58 | 0.50 |
| Cucumbers | 0.33 | 0.25 |
| Dbl Crop Oats/Corn | 0.22 | - |
| Dbl Crop WinWht/Corn | 0.17 | - |
| Dbl Crop WinWht/Cotton | 0.31 | - |
| Dbl Crop WinWht/Sorghum | 0.26 | - |
| Dbl Crop WinWht/Soybeans | 0.12 | - |
| Dry Beans | 0.17 | 0.32 |
| Durum Wheat | 0.23 | 0.20 |
| Fallow/Idle Cropland | 0.39 | 0.50 |
| Flaxseed | 0.27 | - |
| Garlic | 0.42 | 0.30 |
| Grapes | 0.24 | 0.35 |
| Greens | 0.27 | - |
| Herbs | 0.14 | 0.15 |
| Hops | 0.60 | 0.42 |
| Lentils | 0.29 | 0.32 |
| Lettuce | 0.28 | 0.25 |
| Millet | 0.63 | 0.38 |
| Mustard | 0.40 | 0.38 |
| Oats | 0.27 | 0.20 |
| Onions | 0.27 | 0.30 |
| Oranges | 0.10 | - |
| Peanuts | 0.46 | - |
| Pears | 0.27 | - |
| Peas | 0.17 | 0.32 |
| Pecans | 0.20 | - |
| Pistachios | 0.26 | - |
| Plums | 0.26 | - |
| Pop or Orn Corn | 0.20 | - |
| Potatoes | 0.17 | 0.34 |
| Rice | 0.14 | 0.15 |
| Rye | 0.15 | 0.20 |
| Safflower | 0.25 | 0.20 |
| Sod/Grass Seed | 0.14 | |
| Sorghum | 0.37 | 0.38 |
| Soybean | 0.14 | 0.20 |
| Speltz | 0.14 | 0.20 |
| Spring Wheat | 0.17 | 0.20 |
| Sugarbeets | 0.14 | 0.34 |
| Sugarcane | 0.06 | - |
| Sunflower | 0.16 | 0.32 |
| Sweet Corn | 0.35 | - |
| Sweet Potatoes | 0.12 | - |
| Tobacco | 0.50 | 0.49 |
| Tomatoes | 0.33 | 0.25 |
| Triticale | 0.30 | 0.20 |
| Walnuts | 0.23 | - |
| Watermelons | 0.11 | 0.25 |
| Winter Wheat | 0.20 | 0.20 |

**Table S3.** General Climate Models (GCMs) used in this study.

| GCM |
| --- |
| ACCESS-ESM1-5 |
| BCC-CSM2-MR |
| CNRM-ESM2-1 |
| CanESM5 |
| EC-Earth3 |
| GFDL-ESM4 |
| GISS-E2-1-G |
| INM-CM5-0 |
| IPSL-CM6A-LR |
| MIROC6 |
| MPI-ESM1-2-HR |
| MRI-ESM2-0 |
| UKESM1-0-LL |

**Table S4.** List of management practice types.

| Management practice | ICP vs. WCP categorization |
| --- | --- |
| Alley Cropping | ICP |
| Contour Buffer Strips | ICP |
| Contour Farming | ICP |
| Contour Orchard/Other Fruit Area | ICP |
| Cross Wind Trap Strips | ICP |
| Field Border | ICP |
| Filter Strip | ICP |
| Grassed Waterway | ICP |
| Hedgerow Planting | ICP |
| Herbaceous Wind Barriers | ICP |
| Irrigation System | ICP |
| Tailwater Recovery | ICP |
| Riparian Forest Buffer | ICP |
| Strip Crop | ICP |
| Terrace | ICP |
| Tree/Shrub Establishment | ICP |
| Windbreak/Shelterbelt Establishment | ICP |
| Bedding | WCP |
| Hillside Ditch | WCP |
| Surface Drainage | WCP |
| Field Ditch | WCP |
| Water and Sediment Control Basin | WCP |

# SI References